\newif\ifjournal\journalfalse
\documentclass[12pt,preprint]{aastex}

\usepackage{graphicx}% Include figure files

\usepackage{bm}% bold math
\renewcommand{\vec}[1]{\boldsymbol{#1}}

\newcommand{\grad}{\nabla}

\newcommand{\pp}[2]{\frac{\partial #1}{\partial #2}}
\def\jgr{{\itshape J. Geophys. Res.} }
\def\apj{{\itshape Astrophys. J.} }

\def\prl{{\itshape Phys. Rev. Lett.} }
\def\pop{{\itshape Phys. Plasmas} }
\def\aap{{\itshape Astron. Astrophys.} }

\def\mnras{{\itshape MNRAS} }

\shorttitle{Relativistic Guide Field Reconnection}
\shortauthors{Zenitani \& Hoshino}

\begin{document}

%\preprint{APS/123-QED}

\title{The Role of the Guide Field
in Relativistic Pair Plasma Reconnection}

\author{S. Zenitani}
%\affiliation{
\affil{
NASA Goddard Space Flight Center, Greenbelt, MD 20771;
zenitani@lssp-mail.gsfc.nasa.gov
}
\author{M. Hoshino}
%\affiliation{
\affil{
Department of Earth and Planetary Science, University of Tokyo,
7-3-1, Hongo, Bunkyo, Tokyo, 113-0033 Japan
}

%\date{\today}
%\date{submitted to \it{Physics of Plasmas}}

\begin{abstract}
We study the role of the guide field
in relativistic magnetic reconnection
in a Harris current sheet of pair ($e^{\pm}$) plasmas,
using linear theories and particle-in-cell (PIC) simulations.
Two-dimensional PIC simulations exhibit
the guide field dependence to the linear instabilities;
the tearing or reconnection modes are relatively insensitive,
while the relativistic drift-kink instability (RDKI),
the fastest mode in a relativistic current sheet,
is stabilized by the guide field.
Particle acceleration in the nonlinear stage is also investigated.
A three-dimensional PIC simulation demonstrates
that the current sheet is unstable to the RDKI,
although a small reconnection occurs in the deformed current sheet.
Another three-dimensional PIC simulation with a guide field
demonstrates a completely different scenario.
Secondary magnetic reconnection is triggered by
nonlinear coupling of oblique instabilities,
which we call the relativistic drift-sausage tearing instability.
Therefore, particle acceleration by relativistic guide field reconnection
occurs in three-dimensional configuration.
Based on the plasma theories,
we discuss an important role of the guide field:
to enable non-thermal particle acceleration by magnetic reconnection.
\end{abstract}

\maketitle

\section{INTRODUCTION}

Magnetic reconnection in collisionless plasmas
is of strong interest in high-energy astrophysical places
such as active galactic nuclei (AGNs; \citet{dimatteo,birk01}),
extragalactic jets \citep{rom92,lb98,larra03},
pulsar winds \citep{michel82,michel94,coro90,lyu01,kirk03},
gamma-ray bursts \citep{dr02,drs02,uz06} and,
soft gamma repeaters \citep{thom95,thom01,lyut03a,lyut06}.
Since it rapidly releases
stored magnetic energy into plasma kinetic energy,
reconnection is considered as a possible underlying mechanism,
to explain particle acceleration or
bursty emission signatures in these sites,
regardless of plasma composition
(pair plasmas or ion-electron plasmas). 

For instance, in the case of the Crab pulsar,
magnetic reconnection is considered 
in the relativistic radial outflow of pair plasmas.
Since the central neutron star is a fast oblique rotator,
its strong magnetic fields ($\sim 10^{12}$ G) are highly striped
so that field lines are almost toroidal inside the flow.
Magnetic reconnection is expected to occur
in field reversal configuration of toroidal fields,
in order to dissipate the magnetic energy \citep{coro90}. 
In the AGN context,
magnetic reconnection is expected
in association with magnetic loops
from a differentially rotating acceleration disk.
In this situation, the field configuration
involves magnetic fields perpendicular to the antiparallel component.
Indeed, reconnection in sheared configuration is
proposed as an acceleration site
of MeV/GeV electrons \citep{lb97,schopper,nodes03},
due to the field-aligned electric field $E_{\parallel}$. 
In the case of soft gamma repeaters,
a giant flare is expected
in ultrastrong magnetic environment ($\sim 10^{15}$ G)
around central neutron stars (magnetars).
Although there are various theoretical models,
it is expected that giant flares involve
magnetic reconnection in the relativistic pair plasma environment.
%Reconnection may accelerate particles or
%eject plasmas in ultrafast velocities.

However, the fundamental mechanism of
relativistic magnetic reconnection,
as well as the conventional non-relativistic counterpart,
is far from being understood. 
From the magnetohydrodynamics (MHD) viewpoint, steady state reconnection models
have been extended to the relativistic regime \citep{bf94b,lyut03b,lyu05},
although they do not fully agree. 
The relativistic resistive MHD simulation demonstrated
mildly-relativistic Petsheck reconnection \citep{naoyuki06}. 
From the kinetic viewpoint,
series of particle-in-cell (PIC) simulations demonstrated
that relativistic magnetic reconnection
is a powerful acceleration engine
\citep{zeni01,zeni05b,zeni07,claus04},
primarily due to the direct particle acceleration
by the reconnection electric field,
which is perpendicular to the reconnecting magnetic field lines. 
In fact, obtained plasma energy distribution is highly non-thermal,
approximated by the power-law distribution with an index of $-1 \sim -3$
\citep{rom92,zeni01,larra03,claus04}. 
Thus, magnetic reconnection seems to be
a favorable source of non-thermal particles 
at the sites of synchrotron radiation \citep{claus04apj}. 
However, it was reported that
the current sheet configuration is unstable to
the relativistic drift kink instability (hereafter the RDKI),
which occurs in the plane perpendicular to the reconnection plane \citep{zeni05a}.
The RDKI grows more rapidly than reconnection,
and it mainly contributes to plasma heating \citep{zeni07}. 
Thus, plasma heating by the RDKI overwrites the reconnection scenario;
nonthermal particle acceleration is unlikely to occur
in antiparallel field configuration
in relativistically hot pair plasmas.

Pair plasma reconnection has also attracted recent attention
as an equal-mass-limit example of ion-electron reconnection. 
For example, there is a long-standing problem regarding the reason
collisionless reconnection occurs ``faster'' than
predicted by the MHD theories.
It has been widely argued that
Hall physics plays an essential role in maintaining fast reconnection
(the GEM reconnection challenge; \citet{birn01,shay01}),
but fast reconnection in non-relativistic pair plasma
shows a counter-evidence
because it does not involve Hall effects \citep{bessho05}. 
Instead, although its physical interpretation needs refinement,
it seems that the off-diagonal part of the pressure tensor \citep{hesse99}
accounts for the electric field for fast reconnection \citep{bessho07,hesse07}.
Furthermore, it has recently been argued that
late-time dynamical structure may regulate fast reconnection \citep{dau07}.

Guide field configuration,
which contains a perpendicular magnetic field
to the antiparallel components,
is an important generalization for
studying the reconnection problem in a shear or twisted configuration.
Reconnection with the uniform guide field has long been studied
for better understanding of the reconnection structure
\citep{drake77,katanuma80,hoshino87,hori97,hesse99,hesse04}
and for applications to solar flares and
to the Earth's magnetopause \citep{shibata95,quest81,sonne81}.
Recent three-dimensional (3D) PIC simulations \citep{scholer03,drake03,ricci03,silin03,prit04}
discussed reduction of the cross-field activities \citep{scholer03,silin03},
detailed reconnection region structure \citep{prit04},
and enhanced electron acceleration \citep{drake03,ricci03}. 
We note that the acceleration mechanism differs
from the one in the antiparallel counterpart
because the reconnection electric field is
no longer perpendicular to magnetic lines;
guide field reconnection involves
a parallel electric field $E_{\parallel}$
to accelerate electrons \citep{ricci03}. 
In relativistic pair plasmas,
\citet{zeni05b} presented the idea that
the guide field stabilizes the RDKI
and that the new oblique instability
triggers secondary magnetic reconnection.  
Nonthermal particle acceleration by magnetic reconnection 
seems to be a likely process in a relativistic current sheet once again,
under the guide field condition.
%Thus, relativistic guide field reconnection is a possible acceleration engine

The purpose of this paper is to
comprehensively discuss the role of the guide field
in relativistic pair plasma reconnection.
Extending our recent papers
(\citet{zeni05b} and \citet{zeni07},
hereafter referred as ZH05b and ZH07, respectively),
we investigate the linear and non-linear development of
a relativistic current sheet in pair plasmas,
by using fully self-consistent kinetic PIC simulations. 
The conclusion of this paper is that
relativistic guide field reconnection is
a favorable process for nonthermal particle acceleration,
from the viewpoint of detailed plasma theory. 
%Secondary magnetic reconnection is triggered
%via nonlinear coupling of oblique instabilities,
%and then it particle acceleration occurs in three-dimensional configuration.

This paper consists of the following sections.
In \S 2 we describe the simulation setup. 
In \S 3 we study
the two-dimensional evolution of the current sheet with the guide field.
We investigate
how the guide field affects the linear instabilities in \S 3.1,
and then
we study
how particles are accelerated in nonlinear guide field reconnection
in \S 3.2.
In \S 4, we study the 3D evolution of
an anti-parallel current sheet. 
In \S 5, we study the 3D evolution of
a current sheet with a guide field.
We introduce new linear instabilities
in the oblique directions in \S 5.1,
and then we discuss the nonlinear development of
the current sheet in \S 5.2,
in relevance to the two-dimensional counterparts. 
Section 6 contains discussion and the summary.

\section{SIMULATION MODEL}

Throughout this study,
we employ a relativistic Harris model as an initial configuration.
Magnetic field and plasma distribution functions are set
in the following way:
\begin{equation}
\vec{B} = B_0 \tanh(z/\lambda) \vec{\hat{x}} + \alpha B_0 \vec{\hat{y}},
\end{equation}
\begin{equation}
f_{s} = \frac{n_0\cosh^{-2}(z/\lambda)}{4\pi m^2cT K_2(mc^2/T)}
\exp\big[
\frac{ -\gamma_s (\varepsilon - \beta_s mc u_y) }{ T }
\big]
+ \frac{n_{bg}}{4\pi m^2cT_{bg} K_2(mc^2/T_{bg})} \exp\big[-\frac{\varepsilon}{T_{bg}} \big],
\end{equation}
where $B_0$ is the antiparallel magnetic fields,
$\alpha$ is the relative amplitude of the guide field,
$n_0$ is the plasma number density of the current sheet in the proper frame,
$T$ is the plasma temperature including the Boltzmann constant in the proper frame,
$m$ is the positron/electron mass,
$c$ is the light speed,
$K_2(x)$ is the modified Bessel function of the second kind,
the subscript $s$ denotes the species
($p$ for positrons and $e$ for electrons),
$c\beta_s$ is the drift speed of the species, and $u$ is the four velocity.
The parameters $T$ and $n_0$ are defined in the proper frame.
$T_{bg}$ and $n_{bg}$ are the temperature and
the number density  of background plasmas.
In this paper, we choose $\beta_p=0.3$, $\beta_e=-0.3$
(relevant Lorentz factor is $\gamma_{\beta}=1.048$),
$T/mc^2=1$, $T_{bg}/T=0.1$ and $n_{bg}/(\gamma_{\beta}n_0) = 0.05$.
Notice that the Harris model with the uniform guide field
exactly satisfies an equilibrium.
We assume that $B_y$ is negative,
but we will observe the same (mirror) results in the positive cases,
because of the mathematical symmetry in pair plasmas. 
We use 3D particle-in-cell (PIC) code
with periodic boundaries in all three directions.
Because of the field reversal,
we set double current sheets in the $z$ direction.
The typical scale of the current sheet $\lambda$ is set to 10 grids.
The size of the simulation box depends on the problem.
Their sizes are presented in Table \ref{table}.
These conditions are same as our recent studies;
runs R3 and D3 are identical to ones in ZH07 and
runs 3D-A and 3D-B are identical to run A and run B in ZH05b.

In many cases, we impose small artificial electric fields around
$(x, y, z) = (0,0,\pm 3\lambda)$ in the very early stage of simulation runs,
in order to ``force'' reconnection around the center of the main simulation domain.
Typical spatial ranges for the artificial fields are
set to $(\Delta x,\Delta z) \sim (\pm 2\lambda,\pm \lambda)$ in two-dimensional cases
and $(\Delta x,\Delta y,\Delta z) \sim (\pm 2\lambda,\pm 2\lambda,\pm \lambda)$
in 3D cases.
This force is often strong enough and then
reconnection process immediately breaks up soon after short linear growth stage.
In Table \ref{table}, the letter F means that
we observe such sudden breakup of forced reconnection.
S* means that we observe quasi-spontaneous evolution,
because the trigger force is not strong enough.
In these cases, it takes some time
until nonlinear reconnection breaks up, and
we observe the linear growth of the instabilities for a while.
In S cases, we set no trigger force.
In two-dimensional runs for the RDKI,
we do not need the trigger force
because the instability quickly grows from thermal noise.

\section{GUIDE FIELD EFFECT TO THE TWO-DIMENSIONAL PROCESSES}
\label{sec:2Dguide}

\subsection{Linear Growth Rates}

First, we investigate the guide field effect on
the two-dimensional reconnection instabilities:
relativistic reconnection in the $x$-$z$ plane and
the RDKI in the $y$-$z$ plane.
We carry out a series of two-dimensional simulations:
five cases of reconnection
($\alpha = B_y/B_0 = 0,-0.5,-1,-1.5,-5$)
and four cases of the RDKI
($B_y/B_0 = 0,-0.25,-0.5,-1$).
Then, we calculate the linear growth rates of
the fastest growing modes
from the perturbed magnetic fields $\delta \vec{B}$ in the neutral plane. 
%calculated from the $\delta B_x$ perturbation in the neutral plane.
In reconnection cases,
it is sometimes difficult to pick up the ``linear'' mode,
because the linear stage is rather short.
It seems that the linear stage takes longer time,
when the initial artificial impact is zero or weak. 
Therefore, for the most difficult case of $B_y/B_0 = -1.5$ (run R3-C),
we also carry out the relevant spontaneous run (run R3-D)
to confirm the linear growth rate. 
We notice that the linear growth rate is
not identical to the energy release rates in the nonlinear stage.
Once the nonlinear reconnection breaks up,
the linear perturbations quickly cascade into
the longer-wavelength perturbations, which grows more explosively,
$2$-$3$ times faster than the linear or quasi-linear modes.

Along with the PIC simulations,
we evaluate the linear growth rates of
the two-dimensional instabilities by using eigenvalue analysis,
in the same way as our previous work (Appendix B in ZH07). 
In this analysis, 
we use the relativistic fluid equations
\begin{equation}\label{eq:fluid}
\frac{\gamma^2_s}{c^2}
(p_s + e_s) 
( \pp{}{t} + \vec{v}_s \cdot \grad )
\vec{v_s}
=
- \grad p_s +
\gamma_s q_s n_s 
\Big( \vec{E} +  \frac{\vec{v}_s}{c} \times \vec{B}  \Big)
- \frac{\vec{v}_s}{c^2} ( \gamma_s q_s n_s \vec{E} \cdot \vec{v}_s + \pp{p_s}{t} ),
\end{equation}
where $p$ is the isotropic plasma pressure 
and $e$ is the internal energy that contains the rest mass energy.
In addition, we use the continuity, Maxwell equations, and adiabatic gas condition
to close equations.
We assume the perturbation for
arbitrary wavevector $\vec{k}=(k_x,k_y)$;
$\delta f \propto \delta f(z) \exp(ik_x x + ik_y y + \omega_i t)$,
where $\omega_i$ is the linear growth rate
(the imaginary part of the wave frequency),
and then
we solve the linearized equations as an eigenvalue problem. 

Let us discuss two-fluid theory in the $x$-$z$ plane
($k_y=0$ or $\partial/\partial y=0)$.
By ignoring the relativistic drift speed effect
(third term of the right-hand side of eq. [\ref{eq:fluid}]),
simple assumptions ($q_p=-q_e=q$,
$\gamma_p=\gamma_e=\gamma$, $p_p=p_e=p$, $e_p=e_e=e$, and $n_p=n_e=n$)
yield the $y$ component of the generalized Ohm's law
\begin{equation}\label{eq:ohm}
\Big( \vec{E} + \frac{\vec{V}}{c} \times \vec{B} \Big)_y
=
\frac{\gamma (p + e) }{2 c^2 q n }
\Big(\pp{}{t}v_{py} - \pp{}{t}v_{ey} \Big),
%+ \Big( \frac{v_{py}^2+v_{ey}^2}{2c^2} E_y
%+ \frac{v_{py}-v_{ey}}{2c^2\gamma q n}\pp{p}{t} \Big),
\end{equation}
where $\vec{V}=(1/2)(\vec{v_p}+\vec{v_e})$ is
the bulk velocity of pair-plasma fluids.
This indicates that
the fluid inertia of electrons and positrons
account for the effective resistivity
to break down the frozen-in condition and to drive the tearing instability.
The Hall terms are not included,
because they vanish in electron-positron plasmas. 
As discussed in the previous works
on the relativistic tearing instability \citep{zelenyi79}
and the conventional tearing instability,
we obtain purely-growing modes
in the wavevector range of $0 < k_x\lambda < 1$.
We employ the maximum growth rates in the above parameter range
as theoretical eigen growth rates of the reconnection mode. 
In the $y$-$z$ plane
($k_x=0$ or $\partial/\partial x=0)$,
\citet{dau99} demonstrated that
two-fluid theory and Vlasov theory are in good agreement
to explain the drift kink instability in ion-electron plasmas,
in long-wavelength range of $k_y\lambda \lesssim 0.7$.
Similarly, we know that 
our two-fluid theory is in agreement with PIC simulations
in long-wavelength rage of the RDKI; $k_y\lambda \lesssim 0.7$
\citep{zeni05a,zeni07}.
Therefore, we employ $k_y\lambda=0.7$ to
estimate the typical eigen growth rates of the RDKI.
We keep in mind that our theory assumes isotropic fluids
and that it does not contain any kinetic effects nor wave-particle interactions.
These approximations will break down in a thin current sheet,
in which gyro radii is much larger,
and so
kinetic-based theories are more favorable for describing instabilities
(e.g. Vlasov analysis by \citet{bri95}).
However, in the scope of the present paper,
two-fluid theory gives us a plausible estimate of the growth rates.

\begin{figure}[htbp]
\begin{center}
\ifjournal
\includegraphics[width={0.7\columnwidth},clip]{f1.eps}
\else
\includegraphics[width={0.7\columnwidth},clip]{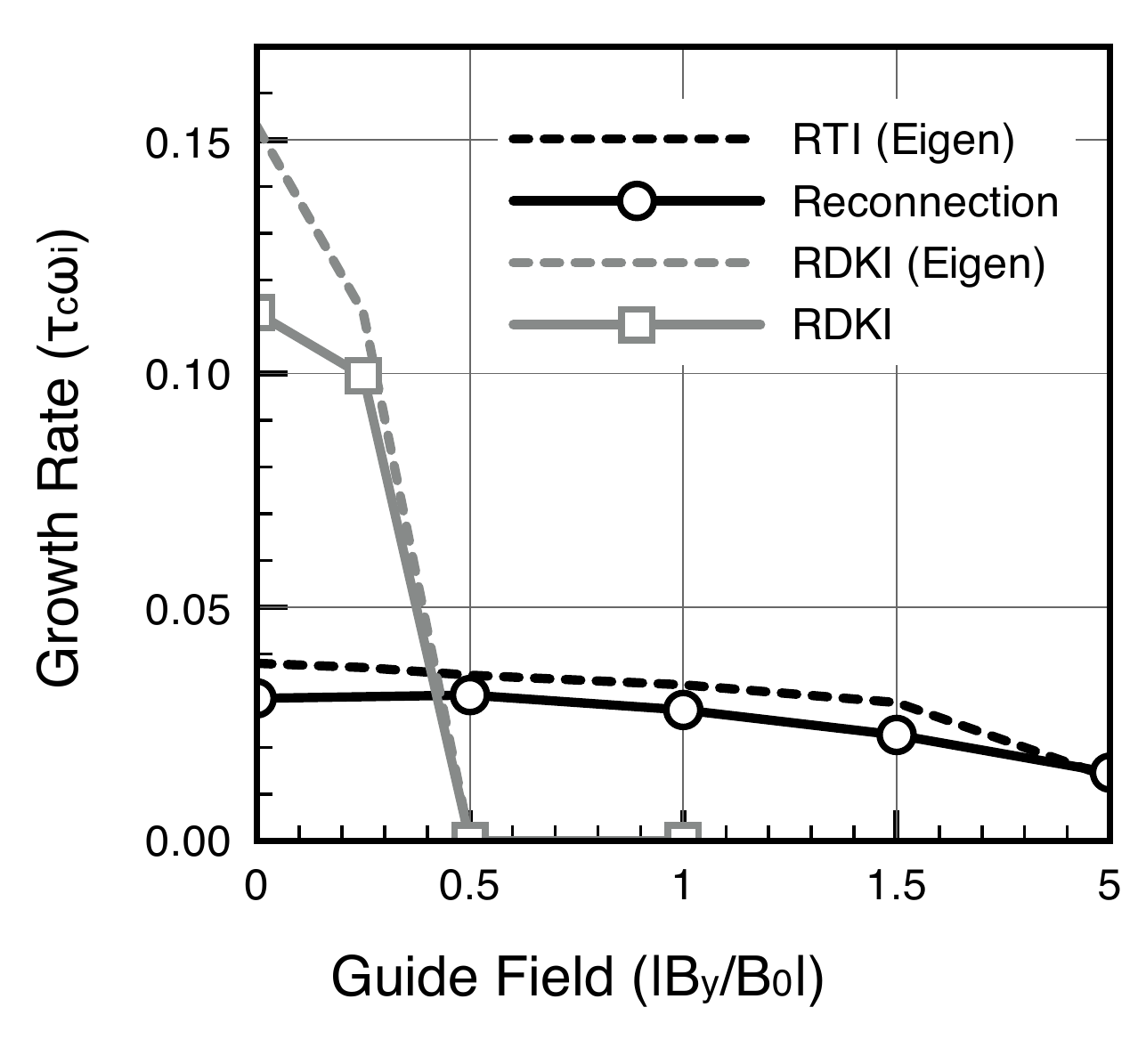}
\fi
\caption{
Guide field dependence ($|B_y/B_0|$) of
the growth rate $\omega_i$,
normalized in the light transit time $\tau_c$.
The fastest eigen growth rate of the relativistic tearing instability,
linear growth rate of the fastest growing modes
in relativistic magnetic reconnection,
eigen growth rate of typical RDKI,
and linear growth rate of the RDKI are shown.
\label{fig:rec_vs_dki_by}}
%\includegraphics[width={0.5\columnwidth},clip]{dki_vs_by.eps}
%\caption{\label{fig:dki_vs_by}
%Guide field dependence ($|B_y/B_0|$) of
%the eigen growth rate ($\tau_c\omega_i$) of
%the relativistic drift kink mode.
%}
\end{center}
\end{figure}

The obtained growth rates of the reconnection mode
are presented in black lines
in Figure \ref{fig:rec_vs_dki_by}
as a function of $|B_y/B_0|$.
In PIC simulations, the fastest reconnection mode
(Fig. \ref{fig:rec_vs_dki_by}, $\textit{black line}$)
has wavenumbers of $k_x\lambda \sim 0.4-0.7$,
which correspond to those
of the relativistic tearing instability.
The maximum eigen growth rates of
the relativistic tearing instability
are presented with the dashed line in Figure \ref{fig:rec_vs_dki_by}.
Both growth rates are in agreement. 
It seems that the growth rates are
relatively insensitive to the guide field.
In the extreme guide field case ($|B_y/B_0|=5.0$),
the growth rate is still the half of that of the antiparallel case.
This tendency is consistent with
reconnection studies in ion-electron plasmas;
an analysis \citep{bri95} and PIC simulations \citep{hori97}
reported that the linear growth rate decreases by a factor of 2 or 3
under strong guide field conditions (up to $|B_y/B_0| = 4$).
It was also reported that
the guide field reduces the quasi-steady reconnection rate
by a factor of $\sim 2$
in PIC simulations and Hall MHD simulations \citep{prit04,huba05}.

The gray lines in Figure \ref{fig:rec_vs_dki_by} show
the growth rates of the RDKI
(gray line by PIC simulation; gray dashed line by the theory). 
The RDKI is rather sensitive to the guide field.
In PIC simulations,
the RDKI spontaneously arises from the thermal noise
before $t/\tau_c=100$
under no/weak guide field conditions
(Runs D3 and D3-A; $|B_y/B_0|=0,0.25$).
However,
we could not observe any sign of the RDKI before $t/\tau_c=200$,
when the guide field is strong enough, $|B_y/B_0| \ge 0.5$.
We also failed to obtain the unstable eigen solutions
when guide field amplitude exceeds a threshold;
$|B_y/B_0| \gtrsim (0.45-0.5)$.
In the physical sense, 
since the wavevector is parallel to the guide field lines,
the magnetic tension of the guide field prevents the RDKI from growing.
Previous ion-electron work \citep{prit04} reported
that the guide field $B_{y}/B_0=0.5$ is
sufficient to stabilize the kink modes.

In summary, in two-dimensional linear regime,
the RDKI grows faster than
the reconnection or the relativistic tearing instability
under no/weak guide field conditions.
However, the situation changes
when we introduce a finite amount of the guide field.
The reconnection is rather insensitive to the guide field,
but the RDKI is suddenly stabilized by the guide field.
Therefore, reconnection will dominate
in the guide field conditions of $|B_{y}/B_0| \gtrsim 0.5$.

\subsection{2D Guide Field Reconnection}
\label{sec:2DPIC}

Next, we study the nonlinear development of
the relativistic guide field reconnection.
Among several simulation runs,
we present the results of run R3-C ($|B_y/B_0|=1.5$),
in which the guide field features are well developed.

\begin{figure}
\begin{center}
\ifjournal
\includegraphics[width={\columnwidth},clip]{f2.eps}
\else
\includegraphics[width={\columnwidth},clip]{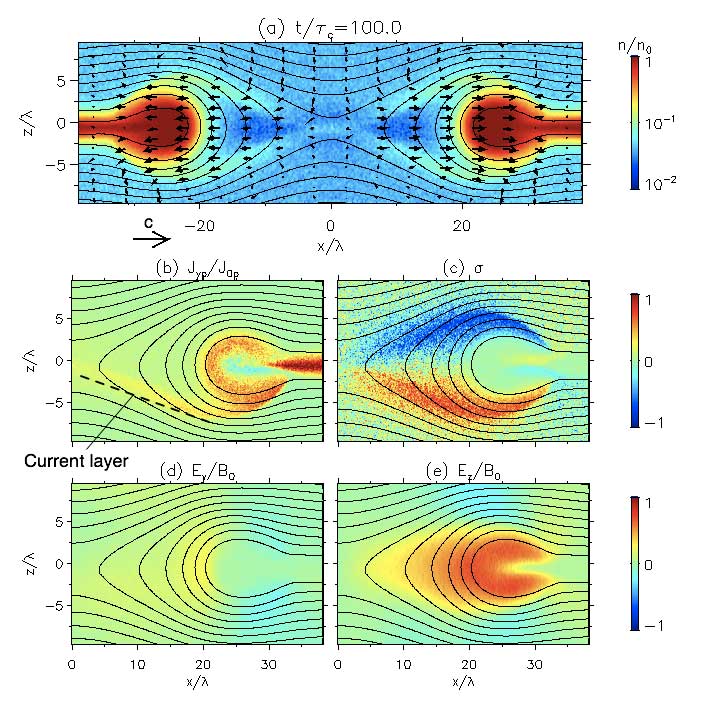}
\fi
\caption{
Snapshots of run R3-C ($|B_y/B_0|=1.5$) at $t/\tau_c=100.0$.
(\textit{a}) Plasma density and flows,
(\textit{b}) positron current density $J_{yp}$,
(\textit{c}) charge non-neutrality [$\sigma=[n_p - n_e]/[n_p + n_e]$],
(\textit{d}) reconnection electric field $E_y$, and
(\textit{e}) vertical electric field $E_z$ are presented, respectively.
The background counter lines show the magnetic field lines in the $x$-$z$ plane.
\label{fig:t100}}
\end{center}
\end{figure}

We force magnetic reconnection around the center
and then reconnection quickly breaks up.
Figure \ref{fig:t100} shows snapshots at $t/\tau_c=100$,
where $\tau_c=\lambda/c$ is the light transit time.
Figure \ref{fig:t100}\textit{a} shows plasma density,
average flow and magnetic field lines.
The maximum outflow speed is $\sim 0.37c$ along the $x$-axis and
it is substantially slower than the antiparallel case of $\sim (0.7$-$0.8)c$.
The composition of plasma flow depends on the quadrants in the reconnection region;
electron flow into the $(+x,-y)$ direction in the $(+x,+z)$ quadrants,
positron flow into the $(-x,+y)$ direction in the $(-x,+z)$ quadrants, 
electron flow into the $(-x,-y)$ direction in the $(+x,-z)$ quadrants
and
positron flow into the $(+x,+y)$ direction in the $(+x,-z)$ quadrants
are dominant. 
The $X$-type layer along the separatrix region is also characteristic.
Figure \ref{fig:t100}\textit{b} shows the $y$-component of the positron current.
The current is strong around the $O$-type region,
while it is weak or reversed in the very center of the $O$-type region,
where the plasma density is at a maximum.
Around the reconnecting region,
there is a broadened structure of weak $y$-current.
The most characteristic thing is
the inclined current layer
along the lower side of the separatrix,
which is indicated by the dashed lines.
This inclined layer corresponds to one of the $X$-type density regions
in Figure \ref{fig:t100}\textit{a}.
The electron current structure is upside down;
it is slightly inclined in the counter-clockwise direction and
it is on the upper side of the separatrix.
In ion-electron plasmas,
several simulations of guide field reconnection reported
an inclined electron current layer \citep{hori97,drake03,hesse04,prit04}
near the $X$-type region.
However, the ion current layer is hardly recognized
due to the ion's large spatial scale.
In our simulation, we observe
both current layers for positrons and electrons.
Positrons are driven by the reconnection electric field $E_y$
around the $X$-type region,
and then
they tend to escape in the lower right (or upper left) direction
along 3D reconnected field lines.
Similarly, electrons tend to escape from the $X$-type region;
they tend to escape in the upper right (or lower left) direction.
Therefore, there is charge non-neutrality
around the outflow region (Fig. \ref{fig:t100}\textit{c}).
Positrons are localized in the lower right side of the reconnection region.
Figure \ref{fig:t100}\textit{d} shows the reconnection electric field $E_y$.
Its typical amplitude is $E_y/B_0 \sim (0.05-0.1)$
around the $X$-type region and
it is substantially weaker than antiparallel case; $E_y/B_0 \sim (0.2-0.3)$.
The $E_x$ component is comparable or slightly larger than $E_y$,
but consistent with plasma inflows.
Its typical value is $E_x/B_0 \sim 0.1$ in the upper inflow region
and $E_x/B_0 \sim -0.1$ in the lower inflow region.
The vertical electric field is
substantially large: $E_z/B_0 \sim 0.7$ in Figure \ref{fig:t100}\textit{e}.
It is consistent with the outflow jet and guide magnetic field $B_y$,
which is compressed around the $O$-type region.
The non-neutral charge distribution is consistent
with the $E_z$ structure, too.

\begin{figure}
\begin{center}
\ifjournal
\includegraphics[width={0.6\columnwidth},clip]{f3.eps}
\else
\includegraphics[width={0.6\columnwidth},clip]{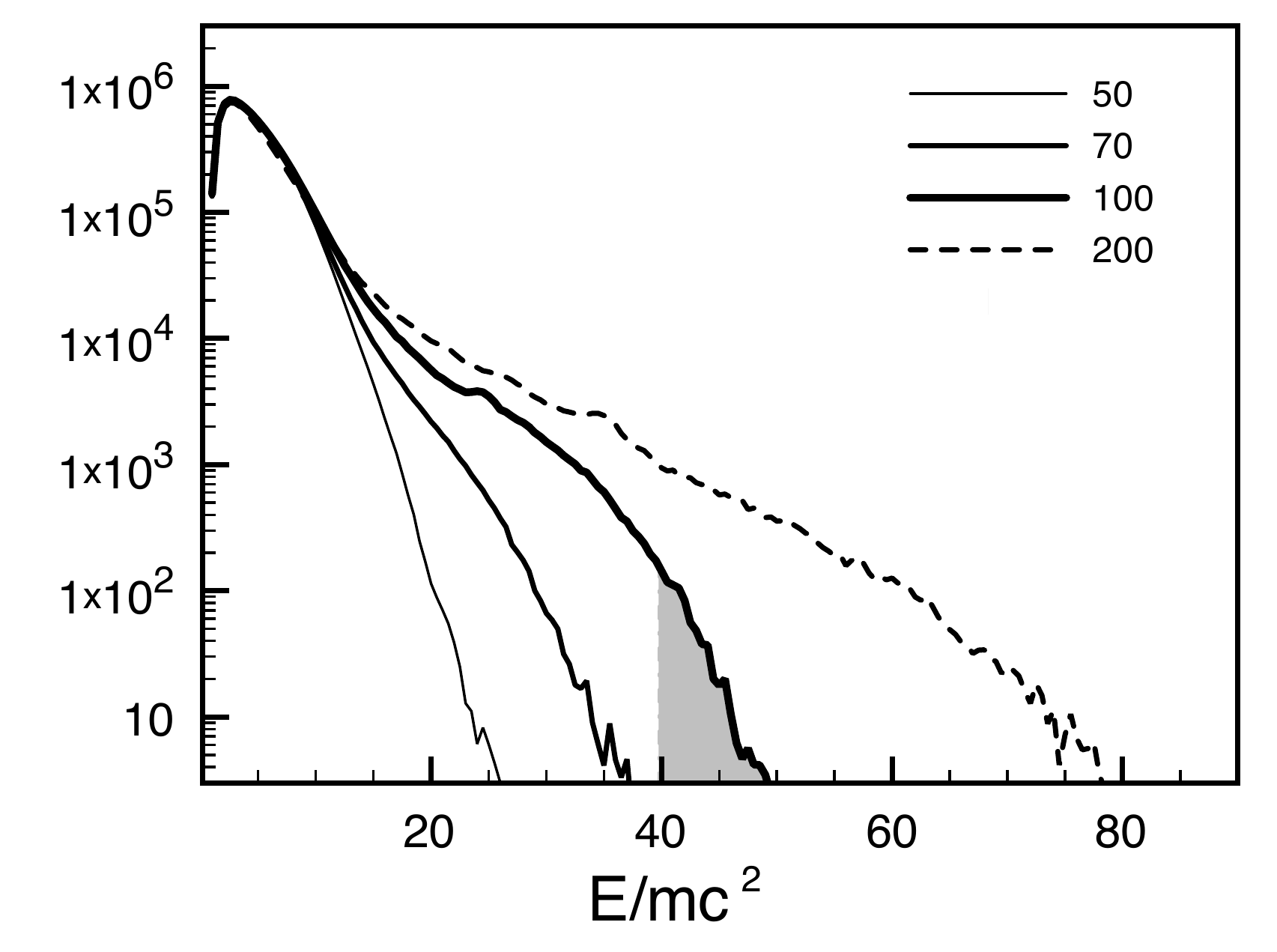}
\fi
\caption{
\label{fig:espec}
Energy spectra of run R3-C at characteristic stages:
$t/\tau_c=50,70,100$, and $200$.}
\end{center}
\begin{center}
\ifjournal
\includegraphics[width={0.95\columnwidth},clip]{f4.eps}
\else
\includegraphics[width={0.95\columnwidth},clip]{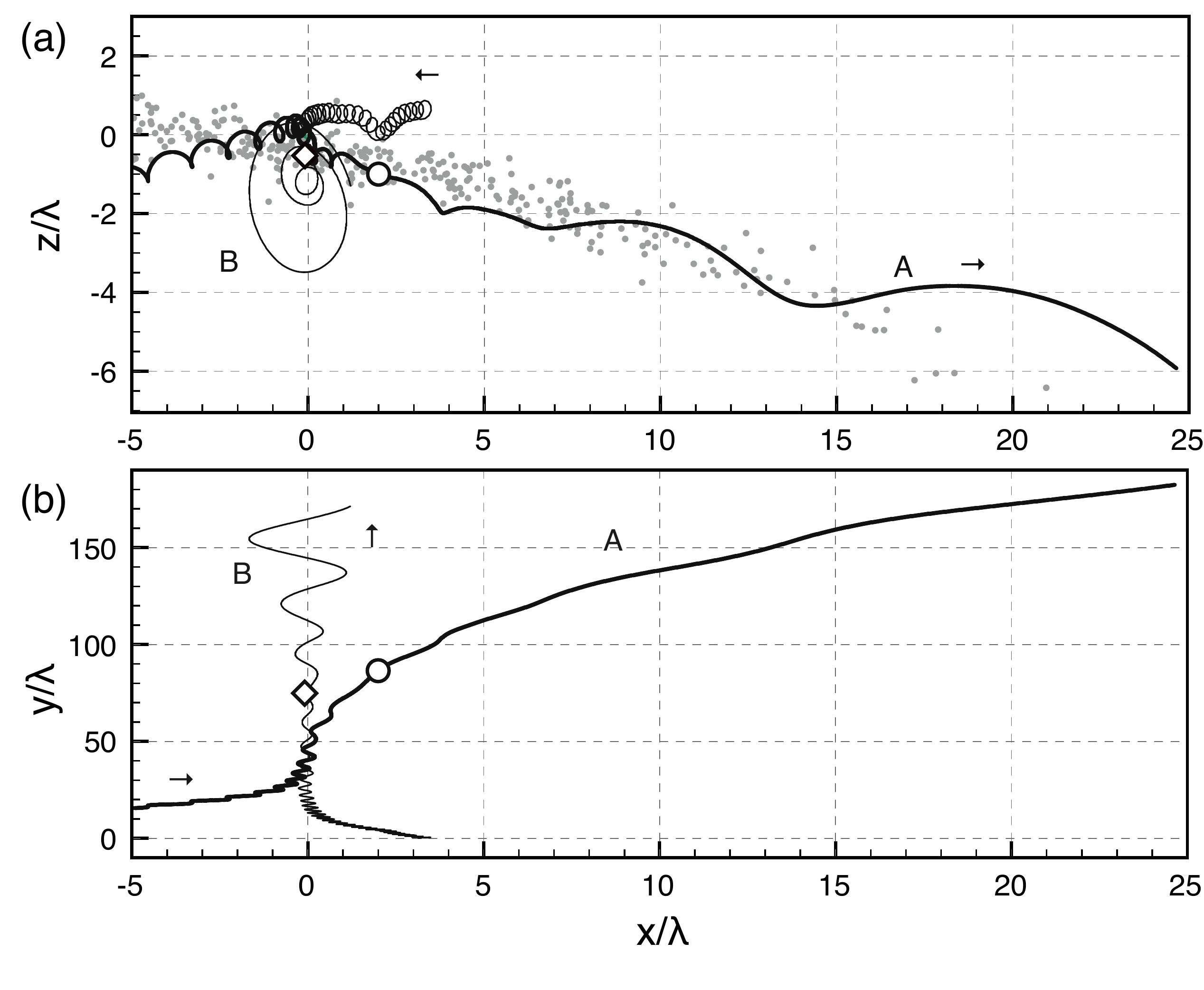}
\fi
\caption{
(\textit{a}) Spatial distribution of high-energy particles
(\textit{gray points}; $\varepsilon \ge 40 mc^2$) at $t/\tau_c=100$.
Two typical trajectories are projected in the same plane.
The signs show the relevant positions at $t/\tau_c=100$.
(\textit{b}) Trajectories in the $x$-$y$ plane.
\label{fig:HE}}
\end{center}
\end{figure}

Figure \ref{fig:espec} shows
the energy spectra over the main simulation domain.
As reconnection evolves, particle acceleration takes place
and then the high-energy tail continues to grow.
The late-time spectrum at $t/\tau_c=200$ is quasi-stable.
Although it is difficult to discuss the spectral index
in such a temporally and spatially limited system,
the power-law index is $\sim -2.9$ in the last stage.

In order to study particle acceleration,
we track the trajectories of the highest energy positrons,
whose energy exceeds $40 mc^2$ at $t/\tau_c=100$
(indicated by gray shading in Fig. \ref{fig:espec}).
Over the entire simulation domain
they are only found along the inclined current layer,
both in the upper left side and
in the lower right side in the reconnection region.
The high-energy electrons are found in the other current layer. 
Gray points in Figure \ref{fig:HE}\textit{a} show
spatial distribution of selected positrons
around the lower right current layer.
We pick up two typical trajectories
from $t/\tau_c=0$ to $t/\tau_c=200$,
which are indicated by solid lines in Figure \ref{fig:HE}.
We call them positron A and positron B, respectively.
The diamond or circle signs show their positions at $t/\tau_c=100$. 
Positron A belongs to the majority.
Starting from the left outside the presented region,
the positron visits the $X$-type region
and then escapes toward the lower right,
gyrating around the reconnected field lines.
The path is along the inclined positron current layer.
While it travels a long distance in the $y$-direction,
the particle mainly gains its energy
from the reconnection electric field $E_y$ near the $X$-type region.
Notice that the scaling of the $y$-direction is
extremely larger than the $x$- and $z$-directions in Figure \ref{fig:HE}\textit{b}. 
The energy gained from $E_x$ and $E_z$ is substantially low. 
%Since particles are threaded by the guide field,
%most of particle energy is stored in $y$-momentum;
%typical $y$-momentum $\langle p_y \rangle/ \langle mc^2 \rangle$ exceeds 20
%in the inclined current layer.
%This is substantially larger than that of the current sheet in the antiparallel case. 
% 
Positron B is a rare sample.
It starts near the center
and then it moves into the $X$-type region.
As reconnection goes on,
it travels in the $y$-direction
along the $X$ line in the $x$-$z$ plane
because it is trapped by the guide field $B_y$ around the $X$ line.
At $t/\tau_c=200$, it still rotates around the $X$ line
and its energy increases from $40.8 mc^2 (t/\tau_c=100)$
to $75.5 mc^2 (t/\tau_c=200)$.
The acceleration continues
until it finds a way to escape toward
the upper left or the lower right.

In the antiparallel reconnection,
the reconnection electric field $E_y$ is a main player
of particle acceleration, too.
Note that the reconnection electric field is
perpendicular to the magnetic field lines
in the antiparallel case, $\vec{E} \cdot \vec{B} \sim 0$.
Therefore, particle acceleration occurs
in the acceleration region
where the frozen-in condition is never satisfied
($|\vec{E}| > |\vec{B}|$; \citet{zeni01}), 
or in the piled-up regions where
the electro-magnetic fields have strong peaks (ZH07). 
In the present case of guide field reconnection,
particle acceleration occurs
because $\vec{E} \cdot \vec{B} \ne 0$ around the reconnection region.
The guide field $B_y$ traps the particles,
preventing them from escaping in the $x$- or $z$-directions,
so that they are accelerated by $E_y$.
The particle acceleration occurs
even when they gyrate around the magnetic field lines,
because it relies on the parallel electric field $E_{\parallel}$.

We have also compared the nonlinear evolutions of
two-dimensional reconnection runs with various guide field amplitudes. 
Generally speaking, as the guide field increases,
signatures of guide field reconnection become more apparent;
electron/positron flows become separated, and then
the inclined current layers and charge separation structure appear.
The reconnection electric field also becomes weaker
because outflow jets becomes slower.
In addition, in spontaneous or nearly spontaneous cases,
it takes a longer time until the nonlinear reconnection process breaks up,
as the stronger guide field is imposed. 

Figure \ref{fig:3d_nonth} compares
the acceleration efficiency in various reconnection runs.
The ratio of the nonthermal kinetic energy to
the total kinetic energy are presented as a function of time.
The ratio is calculated from the energy spectra
in the same way as ZH07 (Appendix A in ZH07);
we assume
an equivalent energy spectra of the relativistic Maxwellian distribution,
which carries the same amount of kinetic energy
as the energy spectra in simulations,
and then we obtain the nonthermal kinetic energy
by subtracting the equivalent thermal spectra
from the simulation spectra in their high-energy tail.
This method may underestimate the amount of the nonthermal energy,
but we can quantitatively compare the acceleration efficiency in a simple way.
%Notice that $\varepsilon_2$ is the
In Figure \ref{fig:3d_nonth},
the simulation time for run R3-A ($|B_y/B_0|=0.5$),
in which the trigger force is slightly weak,
is re-arranged by $\Delta t=35\tau_c$
so that we can directly compare it to the other cases. 
Here we focus on the two-dimensional (2D) reconnection runs;
we discuss the 3D results later again in \S 5.2. 
Roughly speaking,
the acceleration efficiency seems to
decrease as the guide field increases.
The antiparallel run (run R3) is
the most efficient accelerator among the three runs.
The maximum accelerated energy is
also a function of the guide field
during the similar timescale
(See Table \ref{table}).
This is due to its strong reconnection electric field
$E_y \sim 0.3B_0$,
and the reconnection electric field $E_y$ becomes weaker
in the guide field cases.
In run R3-C,
the ``acceleration region'' nearly smears out
due to the weak reconnection field $E_y \sim 0.1B_0$.
On the other hand,
the trapping effect of the guide field
contributes to the acceleration efficiency.
Even though particle acceleration by $E_{\perp}$ becomes less active,
parallel acceleration by $E_{\parallel}$ works instead.
Thus, guide field reconnection (run R3-C) still produces
a substantial amount of nonthermal energy $\sim 15$\%.
Considering the limitation of our simple method,
more kinetic energy will be carried by nonthermal particles. 
In general, guide field reconnection seems to be
an efficient particle accelerator,
which releases more than 15\%-30\% of the plasma energy
into the nonthermal kinetic energy.

%***
%In this stage, multiple reconnecting region appears
%due to the spontaneous growth of the tearing mode,
%and then the accelerated particles
%start to enter the neighboring reconnection regions.
%Probably the presence of multiple reconnection regions
%improves the acceleration efficiency in the moderate guide field case,
%but it is out of the scope of the present paper.
In run R3-A, we notice that
the secondary particle acceleration takes place
in the very late stage (at $t/\tau_c \gtrsim 250$),
and then particles are accelerated up to $\gamma\sim 200$.
The spectral index evolves to a harder value of $\sim -2.2$
in the very late stage (at $t/\tau_c=400$).
Since periodic boundary effects arise in that late stage,
particle acceleration may be enhanced
due to interactions with the multiple reconnection structures.
The secondary acceleration is only found
in the weak guide field case (run R3-A),
and so its long-term evolution needs further investigation. 

When we compare runs R3-D and R3-C ($|B_y/B_0|=1.5$),
the system evolution in run R3-D
delays by approximately $145\tau_c$-$150\tau_c$
in terms of the global energy distribution.
In run R3-D, we observe six tearing islands around $t/\tau_c=200$,
and then they start to collide with each other until
one reconnection region dominates around $t/\tau_c=300$.
The $O$-type islands often have onion-ring density structure,
and the $X$-type current layers connect to the outermost rings.
However, the global energy evolution and
the acceleration efficiency are almost the same.
They seem to be rather insensitive to the initial evolution
whether the system evolves from forced single reconnection
or from multiple tearing islands,
in the moderate guide field cases of $|B_y/B_0|\sim 1.5$.

\section{3D EVOLUTION WITH NO GUIDE FIELD}
\label{sec:3DA}

In this section, we study
the 3D evolution of the current sheet
in antiparallel configuration. 
We look at the simulation results of run 3D-A.
As stated in \S 2,
we imposed small external electric fields around the center,
but the trigger fields are not strong enough to
force reconnection.

\begin{figure}
\begin{center}
\ifjournal
\includegraphics[width={\columnwidth},clip]{f5.eps}
\else
\includegraphics[width={\columnwidth},clip]{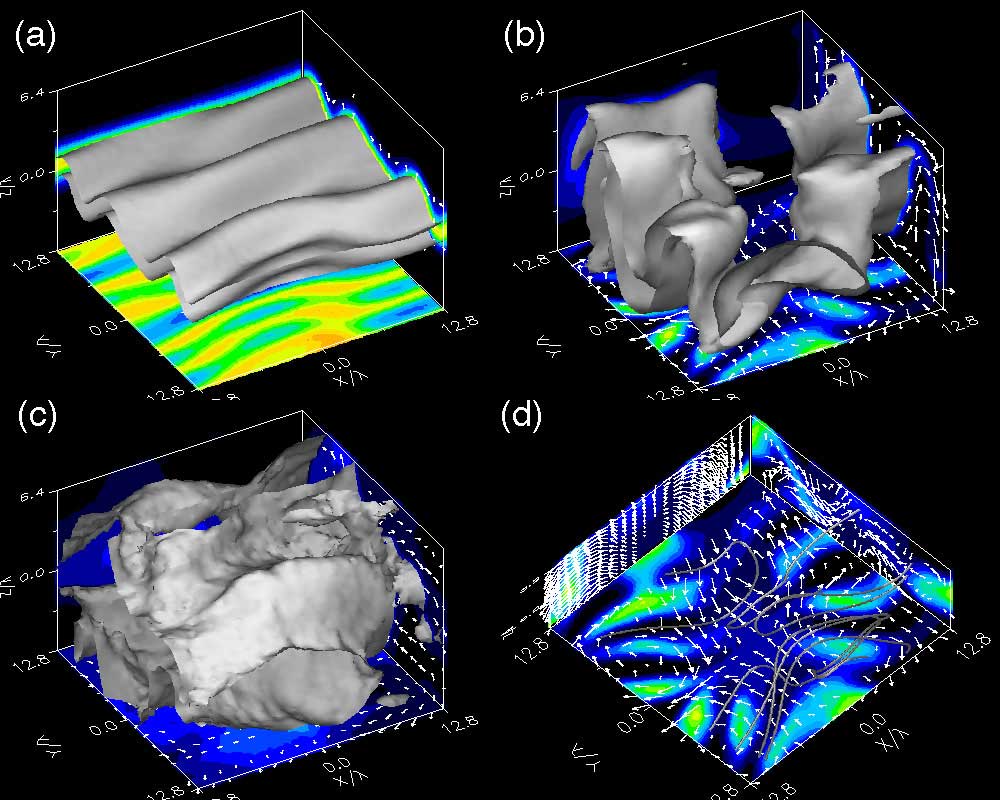}
\fi
\caption{\label{fig:3D-A}
Snapshots of the current sheet in run 3D-A
at (\textit{a}) $t/\tau_c=80$, (\textit{b}) $110$ and (\textit{c}) $140$.
The gray surfaces show the density surface of
(\textit{a}) $n=2/3n_0$, (\textit{b}) $n=1/3n_0$, and (\textit{c}) $n=1/5n_0$, respectively.
The plasma density at the neutral plane ($z=0$) is projected into the bottom wall,
with color from black (empty) through blue (sparse)  to red (dense; $n\sim1.2 n_0$).
Panel (\textit{d}) is a zoomed-in view
around the neutral plane ($-3.2<z/\lambda<3.2$)
at $t/\tau_c=110$.  The gray lines trace the magnetic field lines.
The plasma flow in the 2D planes ($x=0$,$y=0$, and $z=0$)
are presented as white arrows in the three walls.
The light speed ($v=c$) is projected to the arrow length $4$.
\label{fig:3Drec}}
\end{center}
\end{figure}

\begin{figure}
\begin{center}
\ifjournal
\includegraphics[width={0.8\columnwidth},clip]{f6.eps}
\else
\includegraphics[width={0.8\columnwidth},clip]{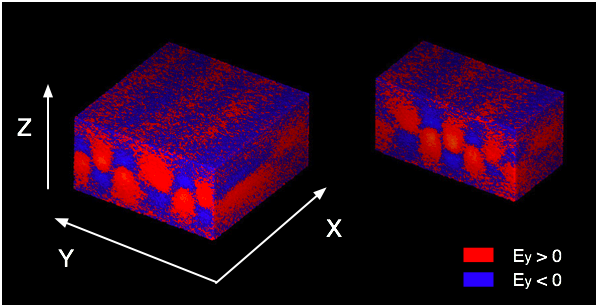}
\fi
\caption{\label{fig:3DEy}
The $E_y$ structure in run 3D-A at $t/\tau_c=80$.
}
\end{center}
\end{figure}

We briefly review the linear evolution,
which was discussed in our previous paper (ZH05b).
Figure \ref{fig:3D-A}\textit{a} is a snapshot at $t/\tau_c=80$.
The current sheet is between the two gray surfaces and
the plasma density at the neutral plane ($z=0$) is
projected into the bottom wall,
with color shading from black (empty) to red (dense; $n=1.2 n_0$). 
Figure \ref{fig:3DEy} presents
the $E_y$ structure at $t/\tau_c=80$.
The red regions have the positive polarity of $E_y > 0$,
while the blue regions are negative: $E_y < 0$.
Apparently these profiles exhibit
a quasi 2D evolution of the RDKI.
One can see the polarity change along the $x$-axis
in the $x$-$z$ plane (the right surfaces in Fig. \ref{fig:3DEy}),
but this is due to small $y$-displacement of the structure.
The wave-number of the RDKI is $k_y\lambda \sim 0.74$ (mode 3),
while mode 4 is observed in the relevant 2D run D3.
However, both mode 3 and mode 4 are reasonable with the linear theories.
The observed growth rate $\tau_c\omega_i=0.06$ is
slightly slower than the expected rate $\tau_c\omega_i\sim0.1$,
but it is still faster than that of the relativistic tearing instability.

Figures \ref{fig:3D-A}\textit{b} and \ref{fig:3D-A}\textit{c}
show the nonlinear development of the current sheets
at $t/\tau_c=110$ and $140$. 
The current sheet is strongly folded at $t/\tau_c=110$,
and then its wave fronts start to collide each other.
At $t/\tau_c=110$,
the gray density surfaces are reset to $n=1/3n_0$,
so that we can see the low-density hole around the center. 
We discover that magnetic reconnection takes place
in this central hole.
Figure \ref{fig:3D-A}\textit{d} is a zoomed-in view
around the neutral plane at $t/\tau_c=110$. 
The gray lines are magnetic field lines,
traced from the following six start points:
$(x,y,z) = (-5,0,-0.5),(-5,2,-0.5) \cdots (-5,10,-0.5)$ in units of $\lambda$.
Magnetic field lines are reconnected around the central low-density region.
Contrary to the 2D reconnection in the $x$-$z$ plane,
reconnection structure involves the $x$-$y$ plane,
because the current sheet is highly folded by the RDKI.
Plasma inflows mainly come from the $\pm y$-directions,
and then
outflows flow into the $\pm x$-direction.
Inflows from the $\pm z$-direction are difficult to identify.
Interestingly, the central reconnection region contains
multiple field reversals between the folded current sheets.
Since plasmas are expelled away from the reconnection region,
there is a low-density hole around the central region.
The reason reconnection takes place in the center
may be the initial trigger impact.
Around the central region,
the RDKI is invoked earlier than other locations, 
%probably due to the triggering perturbation,
and then the folded structure first appears there.
Reconnection occurs in such a well-developed folded region. 
In this simulation, the central reconnection region is
finally overwhelmed by the current sheet corruption,
and then the system evolves into the turbulent state
at $t/\tau_c=140$.
This is consistent with the 2D RDKI picture.
The $x$-$y$ cross sections of the plasma density structure in the upper wall
indicate the plasma mixing across the double periodic boundaries.
Eventually, most energy in the system is
converted into plasma heat,
2 times hotter than the original state.
Particle acceleration by the RDKI or reconnection is negligible.

\section{3D EVOLUTION WITH GUIDE FIELD}

Next, we study a 3D evolution of
the current sheet with a guide field in run 3D-B.
The guide field amplitude is $|B_y/B_0|=0.5$.
All other conditions are the same as
those of run 3D-A in \S \ref{sec:3DA}. 
Throughout the simulation run ($0 \le t/\tau_c \le 220$),
the total energy is conserved within an error of 0.6\%. 
%Snapshots at four different stages
%are presented in the panels in Figure \ref{fig:3D-B}:
%(\textit{a}) $t/\tau_c=120$, (\textit{b}) $170$,
%(\textit{c}) $200$ and (\textit{d}) $220$]
%Gray surfaces show the plasma density.

\subsection{Linear Evolution: Relativistic Drift Sausage/Kink Tearing Instability}

First, we study the linear structure of the mode in detail.
Compared with run 3D-A,
the current sheet seems to be more stable in the early stage.
It takes $t/\tau_c \sim 100$
until we observe visible changes in the current sheet.
At $t/\tau_c=120$, we observe a purely growing flute like mode
in the oblique direction on the upper side of the current sheet
in Figure \ref{fig:3D-B}\textit{a}.
We think
this is a generalized mode between
the relativistic tearing instability and
the RDKI-type instabilities (ZH05b). 
The mode is $(1,1)$ or its wavevector is $\vec{k}_1 \lambda = (0.25,0.25)$. 
The two panels in Figure \ref{fig:cut} show
the plasma density slices at $z=\pm\lambda$.
We see the structure with $\vec{k}_1$
on the upper side of the current sheet at $z=\lambda$,
while we observe another oblique mode
$(1,-1): \vec{k}_2 \lambda= (0.25,-0.25)$
on the lower side of the current sheet at $z=-\lambda$.
On both sides,
it seems that the oblique lines are slightly disconnected
around $(x,y)\sim(\pm 12.8\lambda, 0)$.
Probably the fastest growing modes do not match the system size,
and then they adjust into the nearest periodic mode $(1,\pm 1)$.
Therefore, it takes a longer time before we observe the visible change.
The most powerful Fourier modes are $(1,\pm1)$ in the simulation data,
and Figure \ref{fig:sim} shows $z$-profiles of
the density perturbation of the two modes. 
The amplitude of the two modes is nearly same,
but their peaks are in the other side of the current sheet.
We confirmed that these oblique perturbations are
purely growing unstable modes.
Carrying out a supplemental 2D PIC simulation
under the same conditions in that particular angle, 
% (Run 2D-C in Tab. \ref{table}),
we observe similar a similar asymmetric profile, mode $(2,\pm2)$,
while we obtain mode $(1,\pm 1)$ in run 3D-B.

In order to study the nature of the instability,
we have also calculated eigen profiles of
the oblique instability in the current sheet
by using the relativistic two-fluid theory.
We assume the perturbation of
$\delta f \propto f(z) \exp(ik_x x + ik_y y + \omega_i t)$
as introduced in \S 3.1, and then
we consider the oblique case of $k_x,k_y \ne 0$.
In Figure \ref{fig:rdkti_eigen},
the eigen functions for $k_x \lambda = 0.25$,
$k_y \lambda = 0.25$ and $B_y/B_0 = -0.5$
are presented as functions of $z$.
Perturbed magnetic field profiles ($\delta B_{x}, \delta B_{y}, i\delta B_{z}$)
and electric field profiles ($\delta E_{x}, \delta E_{y}, i\delta E_{z}$) employ
the same normalization.
The other three panels display
the current profiles ($\delta J_{x}, \delta J_{y}, i\delta J_{z}$),
the bulk velocity perturbations ($i\delta V_{+x}, i\delta V_{+y}, \delta V_{+z}$,
where $\delta \vec{V}_{+} = \delta \vec{v}_{p} + \delta \vec{v}_{e}$),
and density profiles
($\delta D_{+} = \delta d_{p} + \delta d_{e}$ and $\delta D_{-} = \delta d_{p} - \delta d_{e}$)
are normalized by their maximum values.
The current profiles ($\delta J_{x}, \delta J_{y}, i\delta J_{z}$)
are calculated from other perturbed values: $\delta D_{+}, \delta\vec{V}_{-}$,
where $\delta \vec{V}_{-} = \delta \vec{v}_{p} - \delta \vec{v}_{e}$.
These profiles are consistent with the simulation data.
For example, the density perturbation $\delta D_{+}$
in Figure \ref{fig:rdkti_eigen}
is in excellent agreement with the relevant perturbation
in simulation data in Figure \ref{fig:sim}.
In general, the perturbation profiles are not
symmetric nor anti-symmetric with $z$.
As observed in $\delta B_x$ or $\delta D_{+}$,
the mode seems to be substantially localized in the upper half of $z > 0$.
The other mode ($\vec{k}_2$; $k_x \lambda = 0.25$, $k_y \lambda = -0.25$)
has an opposite structure;
it is localized in the lower half of $z < 0$.

Figure \ref{fig:illust} schematically illustrates
the structure of the oblique mode ($\vec{k}_1$).
Although we called this mode
relativistic drift-kink tearing instability (RDKTI) in ZH05b, 
we find that
the structure is rather similar to
that of the relativistic drift sausage instability (RDSI),
a cousin mode of the RDKI.
Therefore, we shall rename it
relativistic drift-sausage tearing instability (RDSTI).
For simplicity, we assume that the perturbed structure
is localized in the upper half of the current sheet in the top panel
in Figure \ref{fig:illust}. 
In addition,
the two conventional instabilities in antiparallel configuration;
the RDSI/RDKI and the relativistic tearing instability
are illustrated in the bottom panels
of Figure \ref{fig:illust}. 
The small arrows represent the perturbed vectors.
The phase of the linear perturbation is set to zero at the origin.
Therefore, the dashed line in the neutral sheet is phase $\pi$ and
the imaginary perturbations (phases ${\pi}/{2}$ and ${3\pi}/{2}$)
are found between the origin and the dashed line.
The dashed line (phase $\pi$) in the neutral plane
is equivalent to the $X$-line of the tearing mode.
If we look at the eigen profiles
in the upper vicinity of the neutral plane ($z \gtrsim 0$),
plasma bulk inflow into the $X$ line ($-\delta v_z < 0$),
the diverging outflows in $i\delta v_{x,y}$,
the perpendicular magnetic field $i\delta B_z$,
and the current enhancement $J_y$ are
all consistent with the those of the tearing mode.
In addition, roughly speaking,
the density $\delta D_+$ is positive
in the $O$-like region and
negative in the $X$-like region.
The reconnection electric field $E_y$ is positive around the $X$ line,
although it not strong enough to penetrate into the lower half domain.
There is also $E_z$ structure in $i\delta E_z$,
which is between the $X$- and $O$-type regions.
The $E_z$ structure and the relevant charge separation in $i\delta D_-$
are signatures of the guide field tearing mode or guide field reconnection,
as presented in Figures \ref{fig:t100}\textit{c} and \ref{fig:t100}\textit{e}.

Next, we compare the perturbed structure with
that of the RDSI/RDKI in the $y$-$z$ plane.
We compare the eigen profiles with
those of the RDKI and the RDSI,
which are presented in Figures 14 and 21 in ZH07. 
In both cases of the RDSTI and the RDSI,
the two-peak structure in density profile $\delta D_{+}$
stands for sausage-type modulation,
while reverse peaks stand for kink-type modulation.
The charge separation and $E_z$ structure of the RDSTI
is well connected to the those of the RDSI. 
One specific feature of the RDSTI is
the perturbed plasma flow structure.
The flow direction changes as a function of $z$,
and then the flow is parallel to
the background magnetic field in the topmost layer.
Overall, the structure of the RDSTI is complicated,
but it is well consistent with the relevant 2D instabilities.
The RDKTI has almost same structure as the RDSTI,
except that perturbation is kink-like
in the lower side of the current sheet in the $y$-$z$ plane.

%In the lower half of the current sheet,
%two 2D structures no longer connect each other,
%primary because $0$th order magnetic field contains
%both reverse ($B_x$) and constant ($B_y$) components.
%This also represents the magnetic tension effects.

We investigate the eigen modes over various parameters:
the guide field amplitude $B_y/B_0=0, -0.5, -1, -1.5$ and
the wavevectors of $(0 \le k_x\lambda \le 1, 0 \le k_y\lambda \le 1)$.
Their fastest growth rates ($\tau_c\omega_i$) are 
presented in contour maps in top panels in Figure \ref{fig:map}.
We take $\Delta k_{x,y}\lambda = 0.05$ so that
$21^{2}$ parameters per map are presented. 
All of the modes are purely growing.
Due to the mathematical symmetry,
we can obtain eigen modes and growth rates for $(k_x, k_y)$
from their counterparts in $(|k_x|, |k_y|)$. 
We repeat the validity of our linear theory again.
As discussed in \S 3.1,
our two-fluid approximation will be valid
only in the long-wavelength range
(e.g. $|k_y\lambda| \lesssim 0.7$ for the RDKI/RDSI).
Therefore, the results in Figure \ref{fig:map}
will be reliable
only when $|k\lambda| \lesssim 0.7$.
The bottom panels in Figure \ref{fig:map}
show the type of the fastest eigen mode.
By comparing the peak-structure in their density profiles,
we classify the obtained fastest eigen modes into
the following three types:
kink type (A in Fig. \ref{fig:map}),
sausage type (B), and neither of them (C).
The tearing mode is classified in sausage-type modes,
because its density perturbation is
symmetric with the neutral plane ($z=0$).
Notice that both
kink-type mode and sausage-type mode coexist
in the same point,
and that we discuss the type of the fastest mode
in the specific parameter range.
It seems that the RDKI/RDKTI is faster
only in the shorter wavelength region along the $k_y$-axis.
In the other region, the RDSI/RDSTI replaces the RDKI/RDKTI.

In the guide field case of $B_y/B_0=-0.5$,
the RDKI/RDSI along the $k_y$-axis is stabilized by
the magnetic tension of the guide field
as discussed in \S 3.
The oblique RDSTI modes,
which are along the background magnetic field lines,
survive instead. 
Some oblique modes grow even faster in the guide field case
because they are weakly stabilized
by the magnetic tension in antiparallel case.
As the guide field becomes stronger, 
the most dominant modes change direction,
in accordance with the background magnetic field lines.
Since the RDKI/RDSI component of the instability
is driven by the $\vec{k}$-aligned component of the current,
the growth rates of the oblique modes decrease.
On the other hand, growth rates of the tearing mode component
seems to be rather insensitive to $|B_y/B_0|$. 
In the strong guide field case of $|B_y/B_0|\gtrsim 1$,
we can no longer classify some oblique modes
to kink-type or sausage-type modes (C in Fig. \ref{fig:map}).
These modes are highly localized on one side of the current sheet,
and their perturbation is very small on the other side of the current sheet.

\subsection{Nonlinear Evolution}

\begin{figure}
\begin{center}
\ifjournal
\includegraphics[width={\columnwidth},clip]{f7.eps}
\else
\includegraphics[width={\columnwidth},clip]{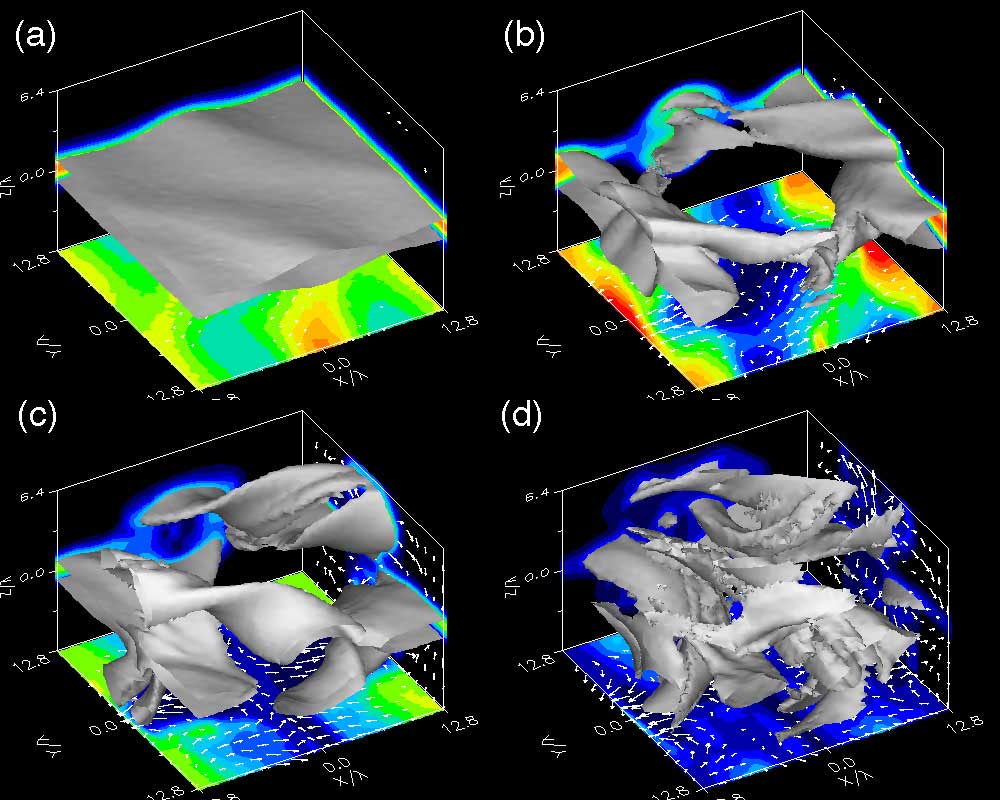}
\fi
\caption{\label{fig:3D-B}
Snapshots of the current sheet in run 3D-B
with a guide field configuration ($B_y/B_0 = - 0.5$) at
(\textit{a}) $t/\tau_c=120$, (\textit{b}) $170$,
(\textit{c}) $200$ and (\textit{d}) $220$.
Gray surfaces show the plasma density of
(\textit{a}) $n=2/3 n_0$, (\textit{b}) $2/3 n_0$, (\textit{c}) $1/2 n_0$ and (\textit{d}) $1/3 n_0$,
respectively.
The bottom walls show the plasma density structure
in/under the neutral plane at (\textit{a-b}) $z=-\lambda$ or
at (\textit{c-d}) $z=0$.
}
\end{center}
\end{figure}

After the linear stage,
the current sheet becomes very thin at several points,
where two RDSTI/RDKTI modes compress
the current sheet from the upper and lower sides.
The thinning point evolves into
a big plasma hole at the center of the simulation box
at $t/\tau_c=170$ in Figure \ref{fig:3D-B}\textit{b}.
The secondary magnetic reconnection takes place there. 
Once reconnection breaks up, it continues to grow.
Figure \ref{fig:3Dcut} shows a 2D slice at $y=0$ at $t/\tau_c=170$.
The upper panel shows the typical reconnection structure.
The outflow velocity is up to $0.6c$,
which is between the antiparallel case ($0.8c$) and
the strong guide field case ($0.4c$) of $B_y/B_0=-1.5$.
The typical inflow velocity is $\sim 0.1c$. 
The current sheet still looks similar to
a Sweet-Parker current sheet, but
careful observation shows
several signatures of the guide field reconnection.
The bottom panel shows the positron current structure in color and the $E_z$ structure in contour, respectively.
The positron current layer is slightly inclined in a clockwise direction.
However, the electron current layer is
inclined in a counter clockwise direction.
So, the two current layers coexist in
a Sweet-Parker-like current sheet.
In addition, the vertical electric field $E_z$ is not negligible.
Its amplitude is $\sim 0.4B_0$ in the right side
and $\sim -0.4B_0$ in the left side,
while the typical reconnection electric field is $E_y\sim 0.17B_0$.
The $x$ component of the inflow electric field is $E_x \sim \pm0.05B_0$. 
These signatures are almost same in the relevant 2D run: run 3D-A.
The width (in $y$) of the reconnection region seems to be limited
by the scale of the system or the scale length of the oblique modes. 
Figure \ref{fig:3D-B}\textit{c} shows a snapshot at $t/\tau_c=200$.
The oblique bridges are blown away from the neutral sheet
due to the intense plasma pressure,
and then the dense points around $(x, y) \sim (0, \pm 12.8 \lambda )$
are no longer observed.
In addition, plasmas are drawn into the central reconnecting point,
and there is a plasma hole along the $X$ line: $x=0,z=0$.
The typical speed of reconnection jets is up to $\sim 0.71c$.
Figure \ref{fig:3D-B}\textit{d} is
the last snapshot of our simulation at $t/\tau_c=220$.
The system structure is highly turbulent,
but we observe the filament-like structures in the $y$-directions.
Considering that reconnection dissipates
the field energy of antiparallel magnetic fields,
it is reasonable that we observe structures
which are threaded by the guide field $B_y$.

After magnetic reconnection occurs,
the magnetic energy tends to be converted
to the nonthermal part of the plasma kinetic energy,
due to the particle acceleration around the reconnection region.
Figure \ref{fig:espec3} compares
the energy spectra in the system
for two 3D runs and the relevant 2D runs of guide field reconnection.
The initial state of two 3D runs is almost similar to
the spectrum of run 3D-B at $t/\tau_c=140$
(Fig. \ref{fig:espec3}, \textit{dotted line}).
In the case of run 3D-A (\textit{bold line}),
plasma energy is converted into plasma heat,
due to magnetic dissipation by the RDKI.
The nonthermal tail of the spectrum ($\varepsilon \gtrsim 30mc^2$)
is slightly enhanced due to the particle acceleration.
The footpoint of the nonthermal tail is approximately
same as that of the 2D RDKI case,
but the tail itself is not as apparent as in the 2D case. 
Due to the irregularity along the $x$-direction or other 3D effects,
particle acceleration by the RDKI
works less effectively than the ideal 2D case. 
In the case of run 3D-B,
the energy spectrum is almost unchanged
until relativistic reconnection breaks up.
At $t/\tau_c = 140$
(Fig. \ref{fig:espec3}, \textit{dotted line}),
plasmas are slightly heated by 3\%
from the initial state.
In the late stage of $t/\tau_c=220$
(Fig. \ref{fig:espec3}, \textit{dash-dotted line}),
the nonthermal tail is enhanced
due to the particle acceleration by reconnection.
The nonthermal slope is well-described
by the power law with the index of $-2.8$
in a range of $8<\varepsilon/mc^2<20$.
This is nearly same as the spectral index in the 2D runs.
The power-law spectral index is $\sim -2.7$ at $t/\tau_c=200$ in run R3-A,
and $\sim -2.9$ at $t/\tau_c=200$ in run R3-C. 

The time history of the nonthermal ratio parameter,
the ratio of plasma nonthermal energy to plasma kinetic energy,
is presented in Figure \ref{fig:3d_nonth}.
In the case of run 3D-A (Fig. \ref{fig:3d_nonth}, \textit{thick line}),
there is a small peak after $t/\tau_c=80$ 
due to the $dc$ acceleration by the RDKI,
but eventually the nonthermal ratio is less than 2\%.
This is consistent with
2D simulations on the RDKI,
which reports the nonthermal ratio of far less than $5\%$ (ZH07).
However, in the case of run 3D-B
(Fig. \ref{fig:3d_nonth}, \textit{thick dashed line}),
shortly after the central reconnection region appears at $t/\tau_c=170$,
more than $14\%$ of the kinetic energy consists of the nonthermal energy. 
This ratio is approximately half of the value
in the relevant 2D run
(run R3-A; Fig. \ref{fig:3d_nonth}, \textit{thin gray line}).
We think this is due to the limited volume of
reconnection region in 3D configuration.
For example, along the neutral line ($X=0$)
at $t/\tau_c=200$ (Fig. \ref{fig:3D-B}c),
reconnection and the relevant particle acceleration is active
around the center $y \sim 0$,
while we do not observe reconnection flow structure
around $y/\lambda \sim \pm 12.8$.
Since reconnection can utilize half of the system volume
along the $y$ direction, the nonthermal ratio is small in run 3D-B,
compared with the 2D counterpart (run R3-A).
If we employ larger simulation box
to reduce the periodic limitation in the $y$-direction,
we may observe wider reconnection region and then
the nonthermal ratio may increase.
Anyway, run 3D-B more efficiently generates
the nonthermal energy than the antiparallel case of run 3D-A,
because underlying physical mechanism is completely different.
Most of magnetic energy is dissipated into
the thermal energy by the RDKI in the anti-parallel case (run 3D-A).
However, in the guide field case (run 3D-B)
a substantial amount of magnetic energy
is dissipated into the nonthermal component of plasma kinetic energy,
associated with magnetic reconnection.

\section{DISCUSSION}

First, we discuss the system evolution in an antiparallel configuration.
We are interested in which process dominates, the reconnection or the RDKI,
because it greatly changes the energy distribution in the system;
reconnection involves nonthermal particle acceleration,
but the RDKI leads to plasma heating.
Comparison of linear growth rates (ZH07) suggested
that the RDKI dominates in the relativistic regime of $T/mc^2 \gtrsim 1$.
Our results in \S 4 basically support this argument.
Although we imposed the external trigger force,
it was not strong enough to evoke reconnection
before the RDKI modulates the current sheet.
The $dc$ acceleration by relativistic reconnection is not likely to evolve;
then magnetic dissipation and plasma heating by the RDKI
would be the main signature of a relativistic current sheet. 
However, we discovered small reconnection regions
inside the folded current sheet structure.
The reconnection generates a density hole around the center,
and then the density hall may evolve into
a global reconnection structure in the larger system.
It is true that
the RDKI grows faster than the reconnection
in our relativistic regime of $T/mc^2 \sim 1$,
but once reconnection is initiated,
we do not know whether it overwhelms the outside RDKI structures.

Next, we discuss the guide field effect on the system evolution.
We showed that the RDKI is completely stabilized by
the finite amount of the guide field in \S 3.
We do not yet understand
how much guide field is necessary to stabilize the RDKI,
but we expect that reconnection mode dominates again
under the guide field conditions.
In \S 5, we showed that the evolution is more complicated than expected.
We considered the oblique instabilities (RDSTI/RDKTI)
to understand the linear evolution of the current sheet.
They grow in two oblique directions (e.g. $\vec{k}_1$ and $\vec{k}_2$),
which can be interpreted as twin extensions of the conventional RDSI/RDKI.
In the antiparallel case
the kink-type mode dominates, and
there were no signatures of sausage-type modes.
In the guide field case,
we found that the sausage-type branch (RDSTI) dominates in run 3D-B.
In some sense this is quite reasonable,
because both the RDSI and the tearing instability modulate the current sheets
in a symmetric way with the neutral plane,
while the RDKI modulates the current sheets in an asymmetrical way.
On the other hand,
since the RDSTI/RDKTI have asymmetric structure
and since they always appear as twins in the guide field cases,
it does not matter whether the oblique modes are sausage-like or kink-like.
The important point is that
the twin oblique modes,
whose wave fronts are parallel to the twisted background magnetic fields,
initiates magnetic reconnection.
Since relativistic guide field reconnection
involves particle acceleration,
the guide field turns on the nonthermal particle acceleration
in the 3D system. 
When we impose a stronger guide field,
one can see that an angle between two RDKTI/RDSTI branches
($\vec{k}_1$ and $\vec{k}_2$)
becomes wider in accordance with the lobe magnetic field lines.
The RDSTI/RDKTI slow down,
but their growth rates are still faster than the relativistic tearing instability
(Fig. \ref{fig:map}).
Under the extreme guide field condition,
since the dominant RDSTI/RDKTI modes are
inclined to the $\pm x$-directions,
coupling between two oblique waves may directly lead to
quasi-2D growth of the tearing instability.

\begin{figure}[htbp]
\begin{center}
\ifjournal
\includegraphics[width={\columnwidth},clip]{f8.eps}
\else
\includegraphics[width={\columnwidth},clip]{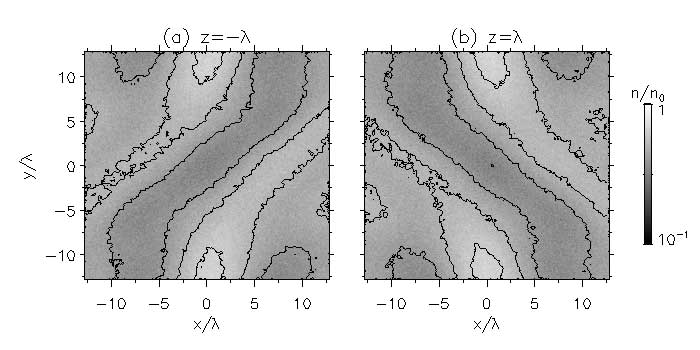}
\fi
\caption{\label{fig:cut}
Slices of the simulation domain in run 3D-B at $t/\tau_c=120$.
The plasma density is represented by gray shading.
}
\ifjournal
\includegraphics[width={0.6\columnwidth},clip]{f9.eps}
\else
\includegraphics[width={0.6\columnwidth},clip]{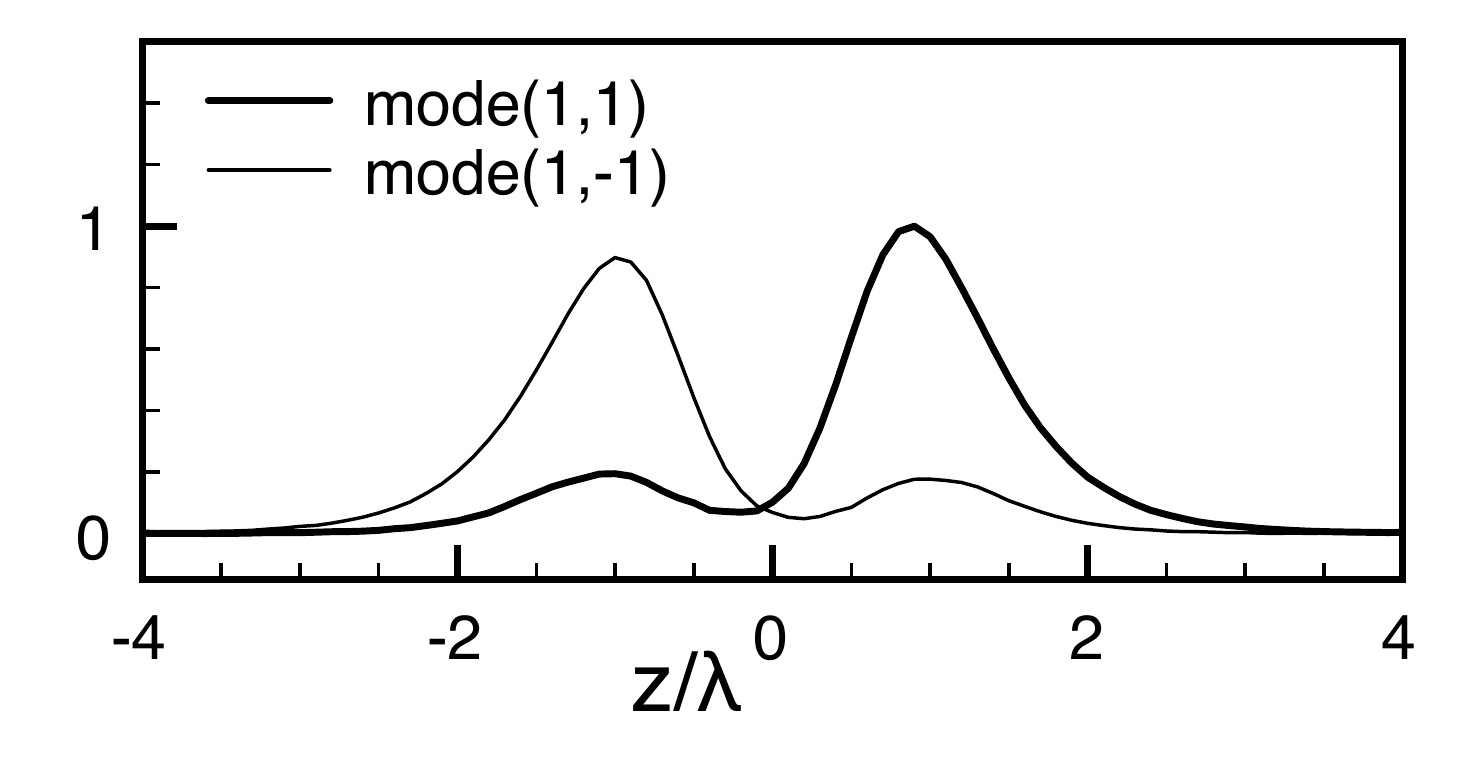}
\fi
\caption{
The $z$-profiles of the density perturbation
in run 3D-B at $t/\tau_c=120$.
Two dominant modes (1,1) and (1,-1) are presented.
\label{fig:sim}}
\end{center}
\end{figure}

\begin{figure}[htbp]
\begin{center}
\ifjournal
\includegraphics[width={\columnwidth},clip]{f10.eps}
\else
\includegraphics[width={\columnwidth},clip]{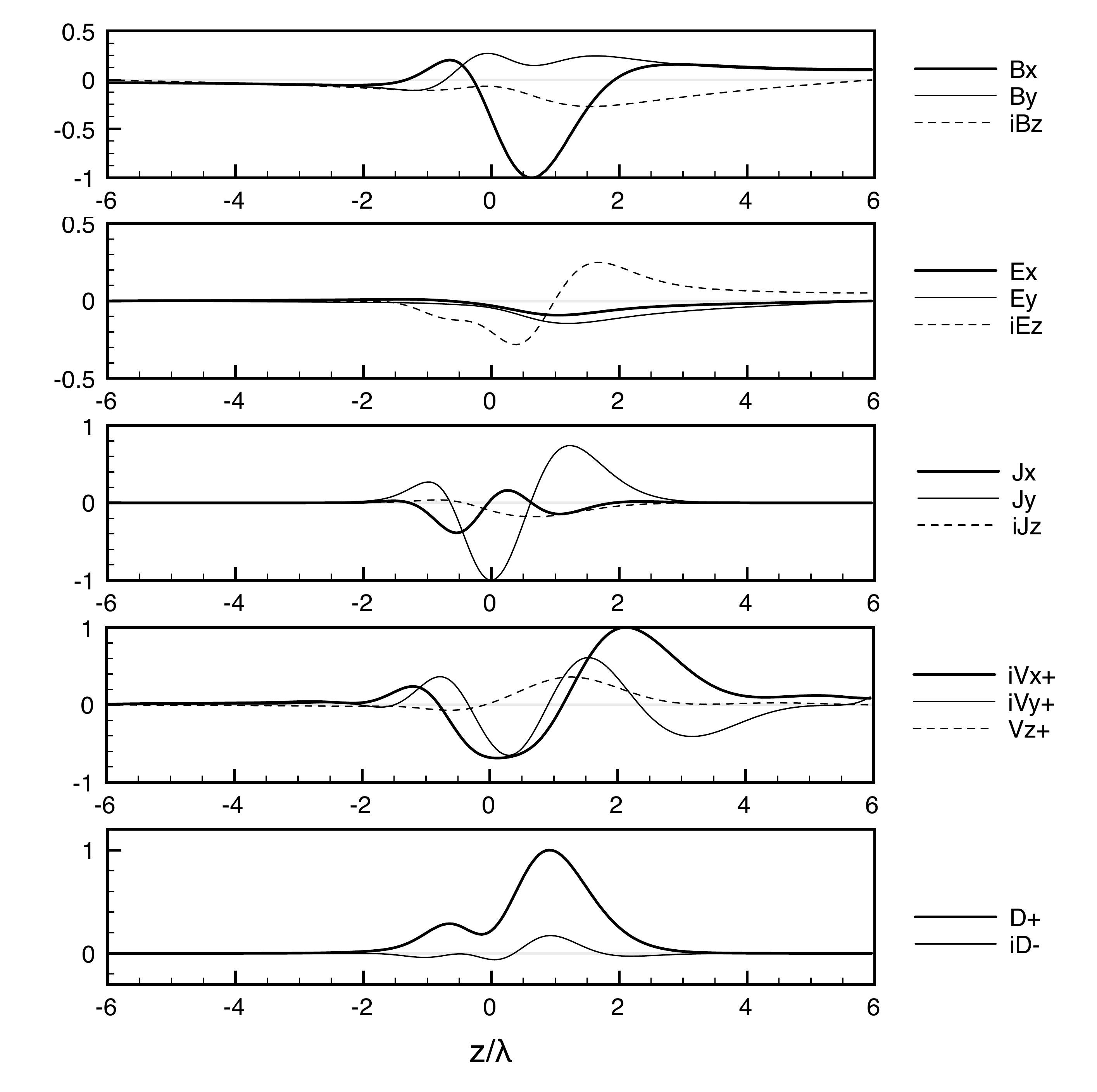}
\fi
\caption{
Typical eigen functions of
the relativistic drift-sausage tearing instability
as a function of $z$,
for the normalized wavenumbers
$k_x \lambda = 0.25$ and $k_y \lambda = 0.25$.
Perturbed magnetic fields: $\delta B_{x}, \delta B_{y}, i\delta B_{z}$;
electric fields: $\delta E_{x}, \delta E_{y}, i\delta E_{z}$;
electric currents: $\delta J_{x}, \delta J_{y}, i\delta J_{z}$;
bulk velocities: $i\delta V_{+x}, i\delta V_{+y}, \delta V_{+z}$;
and
density fluctuations: $\delta D_{\pm} = \delta d_{p} \pm \delta d_{e}$
are presented, respectively.
\label{fig:rdkti_eigen}}
\end{center}
\end{figure}

\begin{figure}[htbp]
\begin{center}
\ifjournal
\includegraphics[width={\columnwidth},clip]{f11.eps}
\else
\includegraphics[width={\columnwidth},clip]{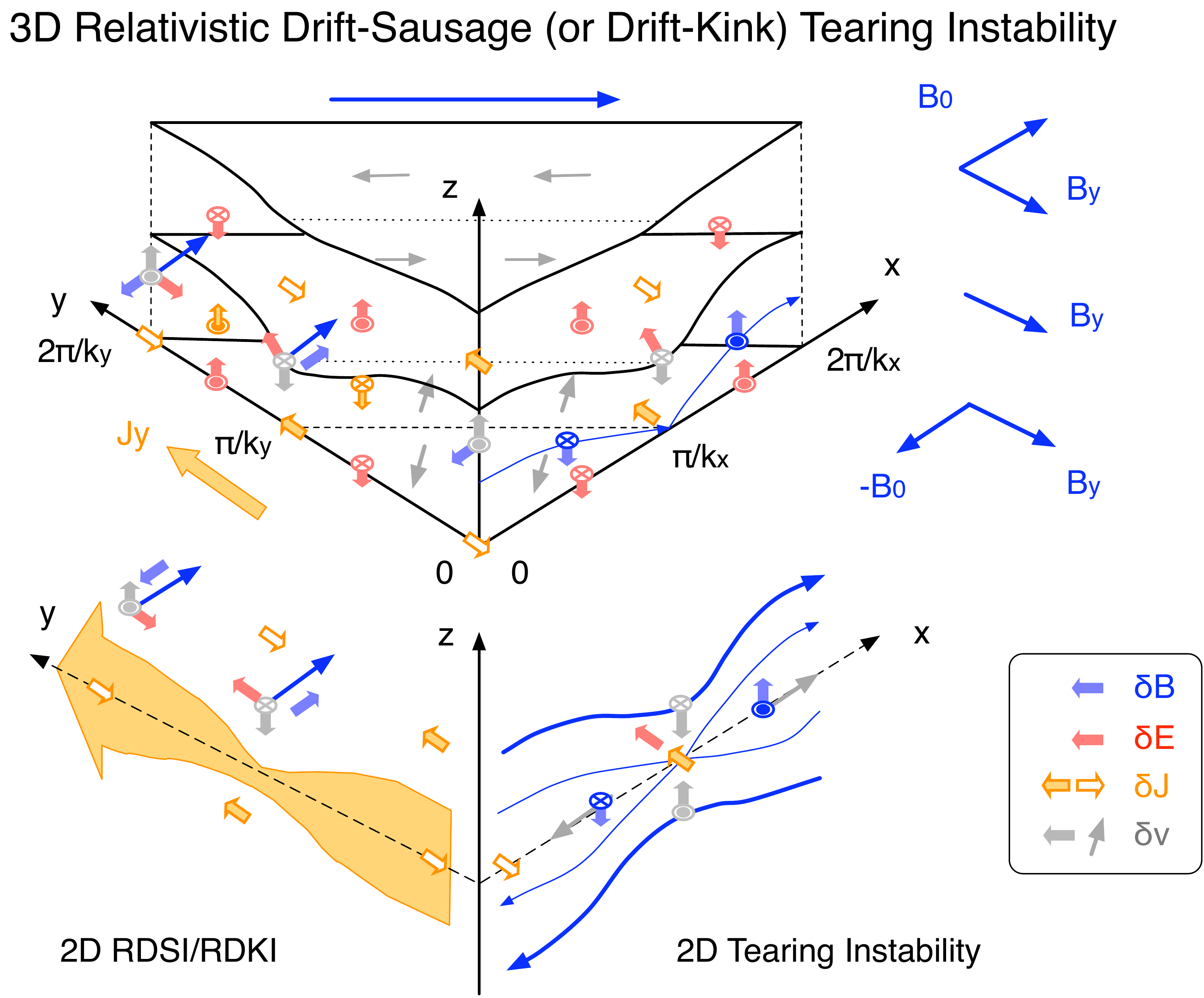}
\fi
\caption{
Schematic illustration of the 3D relativistic drift-sausage tearing instability.
Bottom panels show the relevance to
the conventional 2D instabilities.
\label{fig:illust}}
\end{center}
\end{figure}

\begin{figure}
\begin{center}
\ifjournal
\includegraphics[width={\columnwidth},clip]{f12.eps}
\else
\includegraphics[width={\columnwidth},clip]{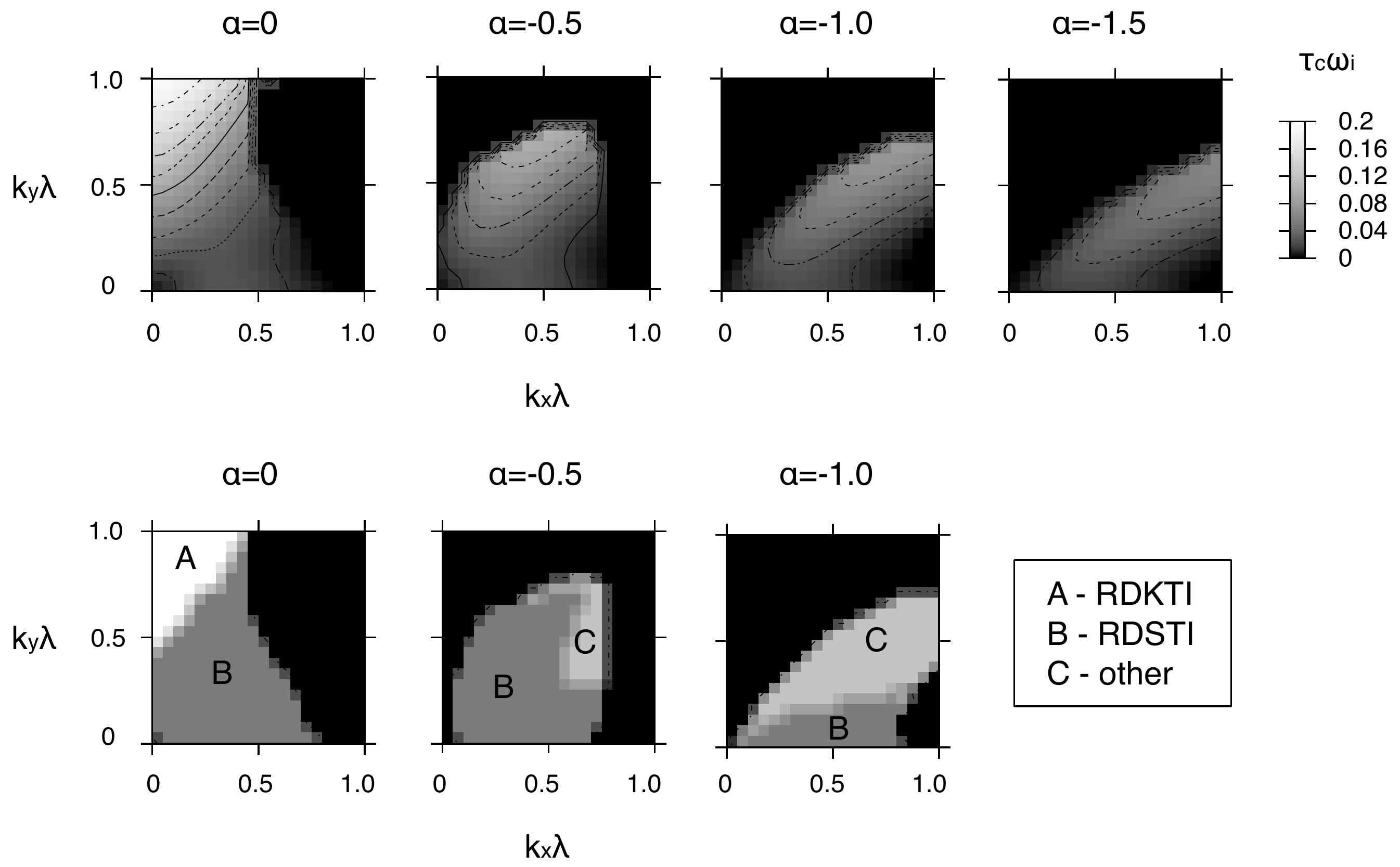}
\fi
\caption{\label{fig:map}
\textit{Top}: Growth rate ($\tau_c\omega_i$) of the instabilities
in wavevector spaces of
$(0\le k_x\lambda \le 1, 0\le k_y\lambda \le 1)$,
as a function of the guide field amplitude $B_y/B_0=\alpha=0,-0.5,-1,-1.5$.
\textit{Bottom}: Three classes of the unstable modes are mapped;
A: kink-type modes; B: sausage-type modes;
and C: others (neither or intermediate).
}
\end{center}
\end{figure}

\begin{figure}
\begin{center}
\ifjournal
\includegraphics[width={\columnwidth},clip]{f13.eps}
\else
\includegraphics[width={\columnwidth},clip]{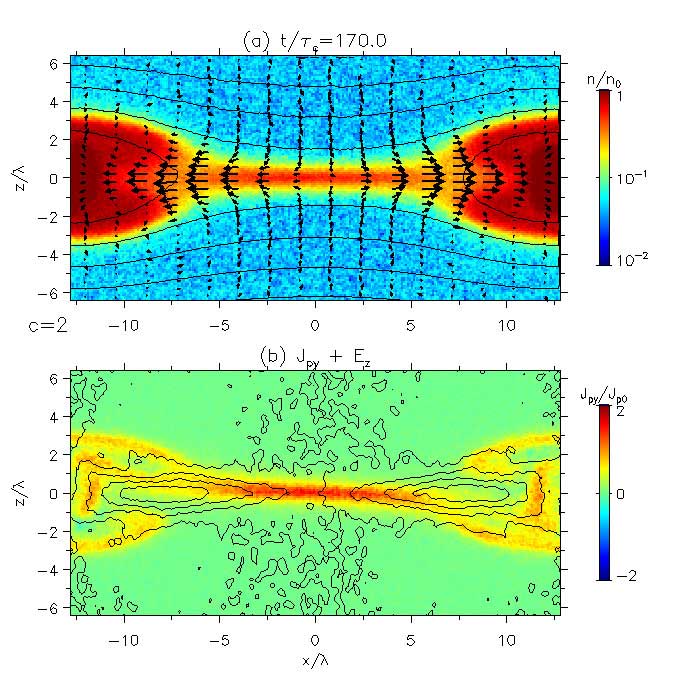}
\fi
\caption{\label{fig:3Dcut}
Snapshots of 2D cut plane at $y=0$ in run 3D-B at $t/\tau_c=170$.
(\textit{a}) Plasma density, flows, 2D magnetic field lines and
(\textit{b}) positron current $J_{py}$ (colored shading),
vertical electric field $E_z$ (contours) are presented.
The contour step $\Delta E_z$ is set to $0.1 B_0$.
}
\end{center}
\end{figure}

\begin{figure}
\begin{center}
\ifjournal
\includegraphics[width={0.65\columnwidth},clip]{f14.eps}
\else
\includegraphics[width={0.65\columnwidth},clip]{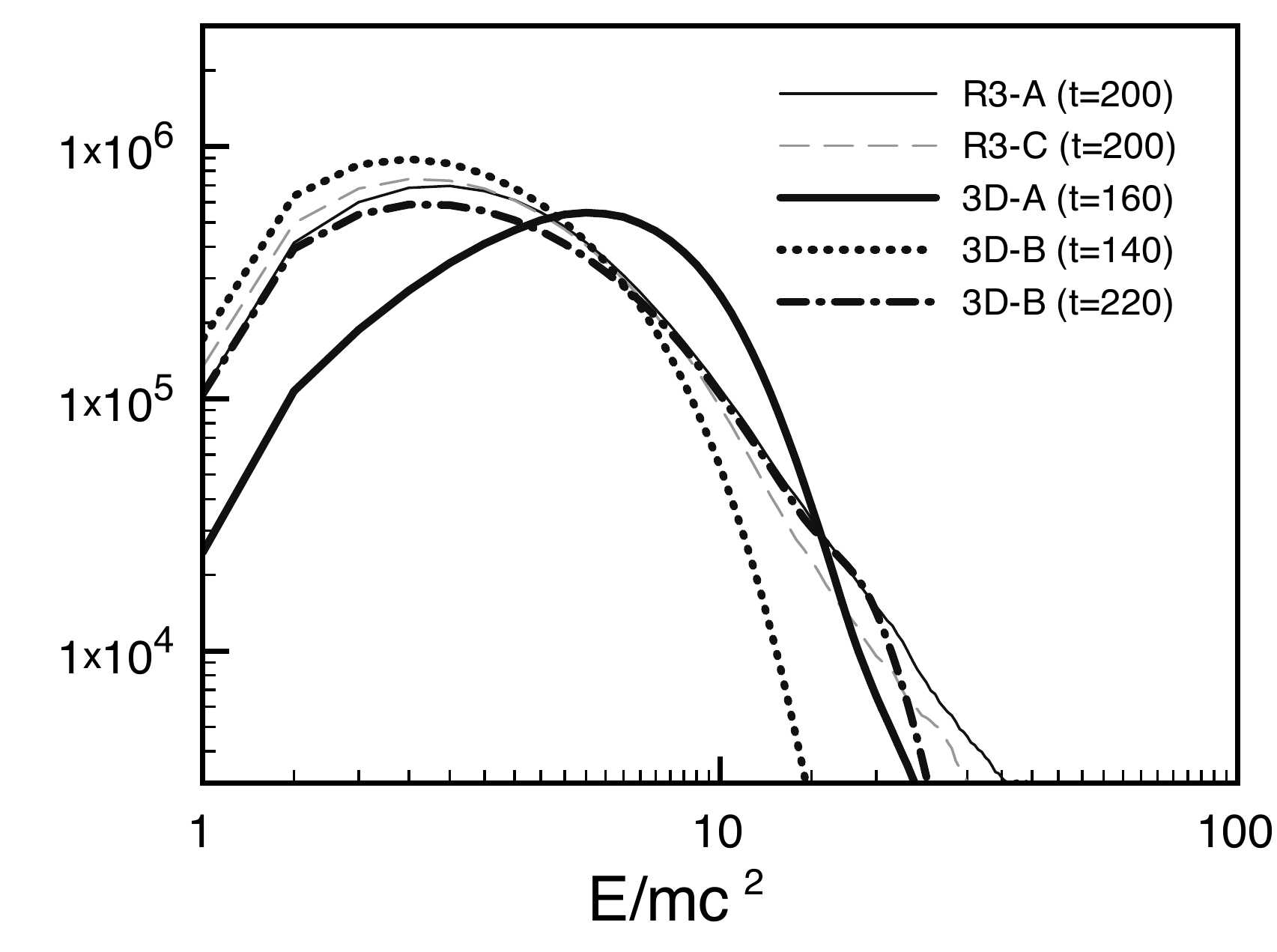}
\fi
\caption{
\label{fig:espec3}
Energy spectra of 3D/2D runs.
Plasma count number for 3D runs are re-arranged for comparison with 2D runs.}
\end{center}
\end{figure}

\begin{figure}[htbp]
\begin{center}
\ifjournal
\includegraphics[width={\columnwidth},clip]{f15.eps}
\else
\includegraphics[width={\columnwidth},clip]{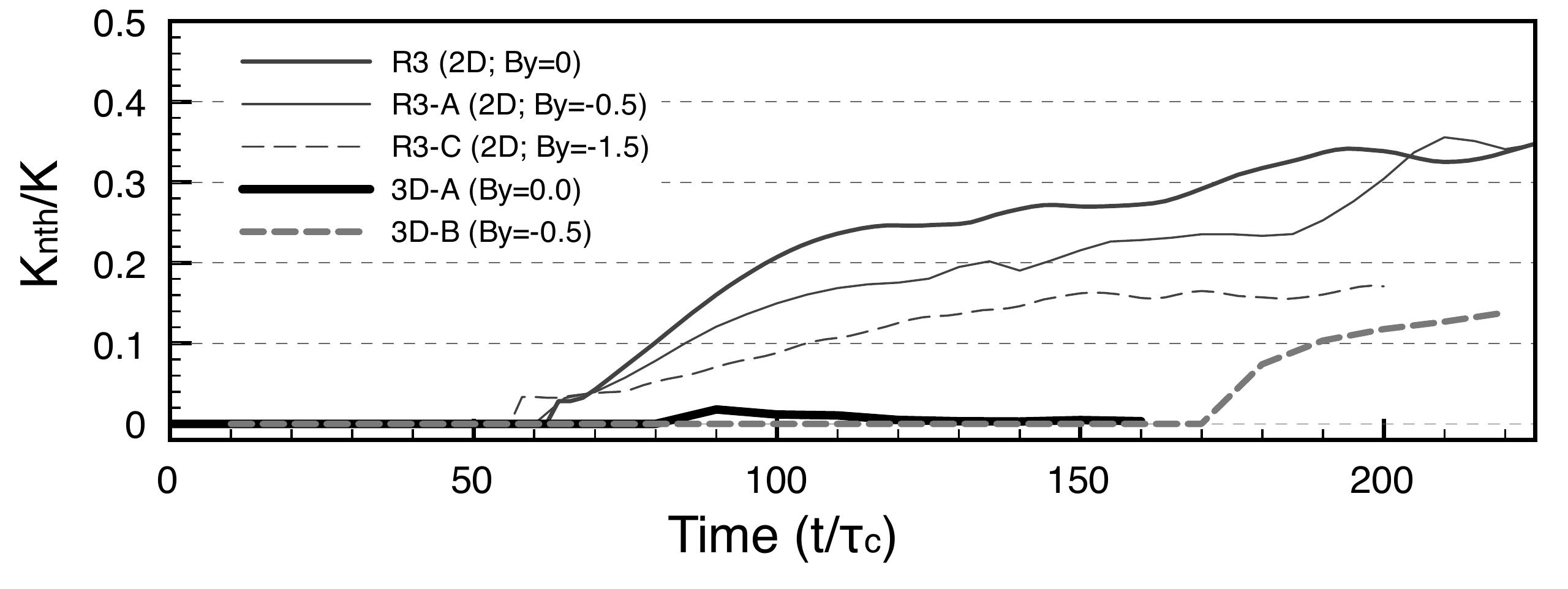}
\fi
\caption{
Time histories of nonthermal ratio parameters for reconnection runs;
runs R3, R3-A, R3-C, 3D-A and 3D-B.
\label{fig:3d_nonth}}
\end{center}
\end{figure}

On the system size limitation,
the late-time evolution of our 3D runs may be somewhat artificial,
due to the periodic boundary effects.
In run 3D-A, magnetic diffusion is enhanced by the plasma transport
across the double periodic boundaries in $z$.
In run 3D-B, the oblique mode (RDSTI) seems to be bounded by the system length.
Since we see the remnant of the other modes (Fig. \ref{fig:cut}),
we probably failed to observe the most unstable modes.
However, as long as similar oblique modes dominate
they will trigger the secondary magnetic reconnection
in the similar way.
If we set larger simulation box,
do we observe multiple reconnection points
in accordance with the spatial structure of the RDSTI/RDKTI?
If so, how are particles accelerated in multiple reconnection regions?
Or does the $X$ line extend and then
does the system evolve into the 2D reconnection?
Furthermore, recent and ongoing works on
large-scale evolution of magnetic reconnection
exhibit more dynamical behaviors than expected
(e.g. plasmoid formation and collisions; \citet{dau07}).
Large-scale, 3D evolution of
relativistic pair plasma reconnection
still remains an open issue,
regardless of the presence of the guide field.
Future simulations may reveal various dynamical behaviours
beyond our linear and early nonlinear results.

Regarding the particle acceleration in the guide field reconnection,
we studied particle acceleration
by the parallel electric field $E_{\parallel}$ in the 2D case,
and we observed similar acceleration signatures in the 3D case, too. 
In general, as long as we investigated
we observed nonthermal energy spectra with indexes of $-2 \sim -3$
by magnetic reconnection, regardless of the amplitude of guide field,
in both 2D and 3D cases.
The upper limit of the accelerated energy remains to be solved,
or it seems unlimited
as long as reconnection continues to consume magnetic energy.
We confirmed that maximum lepton energy easily exceeds
the Lorentz factor of $80$-$200$ ($40$-$100$MeV; see Table \ref{table}).
This will also be influenced by large-scale evolution of the system.

Let us briefly discuss potential astrophysical applications,
based on physical insights from our results.
In pulsar winds with relativistically hot pair plasmas
\citep{coro90,lyu01,kirk03},
it is unlikely that current sheets contain substantial amount of guide fields
because magnetic field lines are highly striped or toroidal.
Therefore, plasma heating by the RDKI is the most likely process,
unless plasma temperature drops down to nonrelativistic one.
On the contrary, in the AGN context,
magnetic reconnection is quite likely
to involve the guide field component.
Several authors demonstrated
electron acceleration due to the field-aligned electric field $E_{\parallel}$
by means of test particle simulations \citep{schopper,nodes03}.
We demonstrated that
relativistic particle acceleration by $E_{\parallel}$
in a fully self-consistent way,
including feedbacks from accelerated electrons and other kinetic effects,
and so our results provide a theoretical proof of
the ultra relativistic particle acceleration in the MeV/GeV range,
at least in electron-positron pair plasma reconnection. 
In the case of soft gamma repeaters,
it is unclear how giant flares occur at magnetars.
The present models discuss
the crustal ``quakes'' \citep{thom95,thom01}
or the flux tube corruption
in analogy with solar flares and coronal mass ejections
to trigger giant flares.
In addition, the tearing instability
in the force-free magnetar corona is proposed to explain
the subsequent bursting activities \citep{lyut03a}.
In these models, magnetic energy is stored in
a magnetic spiral in the star core or the flux lopes,
or coronal magnetic shear. 
Therefore, once a flare occurs,
it is likely that magnetic reconnection involves
a magnetic shear or out-of-plane magnetic field.
The guide field reconnection will occur, and 
reconnection will be a yet another favorable source
of nonthermal particles
as well as ultra-relativistic shock fronts.

Recently, relativistic MHD models have been developed
to investigate astrophysical plasma problems.
Our results warn that an MHD approximation is no longer valid
in the case of relativistic guide field reconnection. 
As plasma outflow becomes an order of $c$,
charge separation structure becomes apparent in the outflow region,
and then it breaks down the charge neutral assumption of one-fluid theory.
Instead, multi-fluid simulations, which deal with
positively charged fluids and negatively charged fluids independently,
are favorable to study with guide field reconnection problems.

Let us summarize this paper.
First, we investigated how the guide field affects 2D instabilities;
the RDKI is stabilized by a finite amount of the guide field,
while the reconnection/tearing mode is rather insensitive to the guide field.
Then, we studied the nonlinear evolution of
relativistic guide field reconnection.
Characteristic field structure and particle acceleration process
were investigated.
Next, we studied 3D evolution of the current sheet.
As predicted by 2D studies,
the RDKI dominates and dissipates the magnetic energy,
but we also discovered that
reconnection occurs inside the folded current sheet.
Finally, we studied 3D evolution
with a guide field condition.
The properties of the oblique RDSTI/RDKTI mode and
the nonlinear evolution of secondary reconnection is discussed,
in association with the 2D counterparts.
Due to the guide field, nonthermal particle acceleration,
which generates power-law energy spectra with an index of $-2 \sim -3$,
occurs in the relativistic pair plasmas.
Our results show that the guide field reconnection is
a favorable acceleration engine in high-energy astrophysical plasmas.

\begin{acknowledgments}
The authors express their gratitude
to T. Yokoyama, M. Hesse, I. Shinohara, P. L. Pritchett and Y. E. Nakagawa
for helpful comments.
The author is also grateful to the anonymous referee
for helping to improve this manuscript.
This work was supported by the facilitates of JAXA
and the Solar-Terrestrial Environment Laboratory, Nagoya University.
\end{acknowledgments}

%\begin{thebibliography}{}
%\end{thebibliography}

% ****** End of file apssamp.tex ******

\clearpage

\clearpage

\clearpage

\clearpage

\clearpage

\clearpage

\clearpage

\clearpage

\clearpage

\clearpage

\clearpage

\clearpage

\clearpage

\clearpage

\begin{deluxetable}{l|ccccccccccccc}
\tabletypesize{\scriptsize}
\rotate
\tablecaption{\label{table} List of Simulation Runs}
\tablewidth{0pt}
\tablehead{
\colhead{Run} &
\colhead{$L_x$} &
\colhead{$L_y$} &
\colhead{$L_z$} &
\colhead{$B_y/B_0$} &
\colhead{${n_{bg}}/{(\gamma_{\beta}n_{0})}$} &
\colhead{$n_{tot}$} &
\colhead{${K_{nth}}/{K}$} &
\colhead{$\tau_c \omega_i$} &
\colhead{$\gamma_{max}$} &
\colhead{Type}
}
\startdata
R3 &
102.4 &
-- &
51.2 &
0 &
5\% &
$1.7 \times 10^{7}$ &
$>3.7 \times 10^{-1} $ &
$3.0 \times 10^{-2}$ &
$> 158 $ &
F &
\\
R3-A &
102.4 &
-- &
51.2 &
-0.5 &
5\% &
$1.7 \times 10^{7}$ &
$>4.0 \times 10^{-1} $ &
$3.1 \times 10^{-2}$ &
$> 122$
($\sim 210$) &
S* &
\\
R3-B &
102.4 &
-- &
51.2 &
-1.0 &
5\% &
$1.7 \times 10^{7}$ &
$> 1.7 \times 10^{-1}$ &
$2.8 \times 10^{-2}$ &
$> 80 $ &
S* &
\\
R3-C &
102.4 &
-- &
51.2 &
-1.5 &
5\% &
$1.7 \times 10^{7}$ &
$1.7 \times 10^{-1}$ &
$2.3 \times 10^{-2}$ &
$> 86 $ &
F &
\\
R3-D &
102.4 &
-- &
51.2 &
-1.5 &
5\% &
$1.7 \times 10^{7}$ &
$>1.8 \times 10^{-1}$ &
$2.3 \times 10^{-2}$ &
$> 79 $ &
S &
\\
R3-E &
102.4 &
-- &
51.2 &
-5.0 &
5\% &
$1.7 \times 10^{7}$ &
- &
$1.5 \times 10^{-2}$ &
$> 42 $ &
S* &
\\
D3 &
-- &
25.6&
51.2 &
0 &
0 &
$4.2 \times 10^{6} $ &
$7.2 \times 10^{-2}$ &
$1.1 \times 10^{-1}$ &
$48 $ &
S &
\\
D3-A &
-- &
25.6&
51.2&
-0.25 &
0 &
$4.2 \times 10^{6} $ &
- &
$1.0 \times 10^{-1}$ &
$38 $ &
S &
\\
D3-B &
-- &
25.6&
51.2&
-0.5 &
0 &
$4.2 \times 10^{6} $ &
- &
N/A &
$22 $ &
S &
\\
D3-C &
-- &
25.6&
51.2&
-1.0 &
0 &
$4.2 \times 10^{6} $ &
- &
N/A &
$22 $ &
S &
\\
3D-A &
25.6 &
25.6 &
25.6 &
0 &
5\% &
$5.4 \times 10^{8} $ &
$1.8 \times 10^{-2}$ &
$6 \times 10^{-2}$ &
$ 52 $ &
S* &
\\
3D-B &
25.6 &
25.6 &
25.6 &
-0.5 &
5\% &
$5.4 \times 10^{8} $ &
$> 1.4 \times 10^{-1}$ &
- &
$> 45 $ &
S* &
%\\
%2D-C &
%36.2 &
%-- &
%25.6 &
%-0.5 &
%5\% &
%$6.9 \times 10^{6} $ &
%- &
%- &
%S &
\enddata
\tablecomments{
Initial parameters and obtained physical values of simulation runs.
R3, D3, and 3D runs are
2D simulations for magnetic reconnection,
2D simulations for the RDKI,
and 3D simulations, respectively. 
%Run 2D-C is a supplemental two dimensional run for 3D-B,
The system width $L_x$ is assumed to be
in the oblique direction ($x$-$y$).
The system size
($L_x,  L_y, L_z$ in units of $\lambda$),
the guide field amplitude ($B_y/B_0$),
the ratio of the background plasma density ($n_{bg}$)
to the plasma density in the current sheet ($\gamma_{\beta}n_{0}$),
the total number of the particles ($n_{tot}$),
the nonthermal ratio ($K_{nth} / K$)
the linear growth rate ($\tau_c \omega_i$)
and
the Lorentz factor of the highest energy particle ($\gamma_{max}$)
are presented.
}
\end{deluxetable}

\end{document}